\documentclass[manuscript]{acmart}
\usepackage{graphicx}
\usepackage{subcaption}
\usepackage{mathtools}
\usepackage{algorithm}
\usepackage{listings}
\usepackage[noend]{algpseudocode}
\usepackage[font=small,labelfont=bf]{caption}
\usepackage{xcolor} 
\definecolor{listinggray}{gray}{0.9}
\definecolor{lbcolor}{rgb}{0.9,0.9,0.9}
\definecolor{Darkgreen}{rgb}{0,0.4,0}
\definecolor{White}{rgb}{1,1,1}
\graphicspath{ {diagrams/} }

\makeatletter
\newcommand\tabcaption{\def\@captype{table}\caption}
\newcommand\figcaption{\def\@captype{figure}\caption}
\def\BState{\State\hskip-\ALG@thistlm}
\makeatother

\newlength{\halfwidth}
\AtBeginDocument{%
  \providecommand\BibTeX{{%
    \normalfont B\kern-0.5em{\scshape i\kern-0.25em b}\kern-0.8em\TeX}}}

\begin{document}

\title{RayNet: A Simulation Platform for Developing Reinforcement Learning-Driven Network Protocols}

\begin{abstract}
Reinforcement Learning (RL) has gained significant momentum in the development of network protocols. However, RL-based protocols are still in their infancy, and substantial research is required to build deployable solutions. Developing a protocol based on RL is a complex and challenging process that involves several model design decisions and requires significant training and evaluation in real and simulated network topologies. Network simulators offer an efficient training environment for RL-based protocols, because they are deterministic and can run in parallel. In this paper, we introduce \textit{RayNet}, a scalable and adaptable simulation platform for the development of RL-based network protocols. RayNet integrates OMNeT++, a fully programmable network simulator, with Ray/RLlib, a scalable training platform for distributed RL. RayNet facilitates the methodical development of RL-based network protocols so that researchers can focus on the problem at hand and not on implementation details of the learning aspect of their research. We developed a simple RL-based congestion control approach as a proof of concept showcasing that RayNet can be a valuable platform for RL-based research in computer networks, enabling scalable training and evaluation. We compared RayNet with \textit{ns3-gym}, a platform with similar objectives to RayNet, and showed that RayNet performs better in terms of how fast agents can collect experience in RL environments.
  
\end{abstract}



\author{Luca Giacomoni}
\email{l.giacomoni@sussex.ac.uk}
\orcid{0000-0003-1535-7558}

\author{Basil Benny}
\email{bb314@sussex.ac.uk}
\orcid{0000-0003-4168-6961}

\author{George Parisis}
\email{g.parisis@sussex.ac.uk}
\orcid{0000-0002-1298-7143}
\affiliation{%
  \institution{University of Sussex}
  \city{Brighton}
  \country{UK}
}

\maketitle

\section{Introduction}
\label{introduction}

Reinforcement Learning (RL) has gained substantial momentum in developing congestion control algorithms \cite{xu2019experience, abbasloo2020classic}, routing \cite{mammeri2019reinforcement}, video rate control \cite{huang2018qarc, mao2017neural}, network access \cite{wang2018deep, naparstek2018deep}, network security \cite{nguyen2019deep, uprety2020reinforcement} and proactive caching \cite{zhu2018deep, he2017integrated}. RL-based algorithms have the potential to adapt to a wide range of network deployments, topologies, traffic workloads and link technologies, without requiring engineering and tuning of bespoke algorithms to deal with said diversity. This, in turn, opens up opportunities for developing future-proof algorithms that can be trained offline, and continuously optimise their behaviour as networks and user workloads evolve. RL-based protocols are still in their infancy and substantial research is required to yield deployable solutions \cite{fuhrer2022implementing}. Developing an RL-based protocol is a complex process that requires (1) deciding on a particular RL algorithm, (2) devising a suitable and effective RL model for the problem at hand, (3) training agents in realistic network setups, and (4) deploying them in the wild. The first two of these involve a range of design decisions related to the action and state space of the RL agent(s), the reward function and the RL algorithm itself. Training the agent(s) is far from trivial; a potentially large number of hyper-parameters (e.g., the discount rate \textit{gamma}, the size of replay buffer, etc.) need to be explored to ensure that the resulting policy is the best possible given the selected RL algorithm and model, the training setup (e.g., in terms of the used network parameters, topology and workload) and the expected deployment parameters. At the same time, training an agent requires the collection of very large amounts of agent experience, which can be done either using a real network (e.g., as in \cite{abbasloo2020classic}), a network emulator (e.g., as in \cite{sacco2021owl}) or network simulations (e.g., as in \cite{tessler2022reinforcement}). 

We argue that network simulators, in particular discrete event, packet level simulators, provide a very effective platform for training RL-based network protocols and respective algorithms. First, simulators offer a \textit{fully controllable and configurable} experimentation environment. Network simulators employ domain-specific languages that make defining networks to be simulated, including the underlying topology, link characteristics, and traffic workloads, easy. Moreover, simulations can run independently and in parallel to each other. It would be extremely expensive and time-consuming to enable such configurability on a real network, and network emulations, such as the ones used in \cite{netravali2015mahimahi}, support limited configurability. Second, training an agent for a network protocol requires exchanging traffic between multiple endpoints. In a real or emulated network, said traffic must be actually played out in the network which, depending on the training setup and parameters, may add a substantial overhead in the training process, in terms of the time it takes to collect agent experience (e.g., in a scenario where the network capacity is very small or when the agent is still far from optimised). To make things worse, training congestion control policies for high-speed networks can be very problematic (or, in fact, impossible) due to the agent needing actions much faster than what the trainer can calculate.\footnote{For example, in \cite{abbasloo2020classic} the authors fix the agent's action calculation/monitoring interval to 20ms, indicating that much lower values would not be possible.} On the contrary, network traffic generated by inefficient policies or traffic that crosses low-bandwidth links can be played very quickly within a simulated network, while the simulated nature of the deployment eliminates the issue around high-performance network deployments. Third, network simulators support \textit{reproducibility of results} by design, which is crucial for RL-based approaches; it is surprising that many papers in RL-based network technologies are being published without any provision for (or even discussion around) reproducibility\footnote{A shining exception is Remy, one of the earliest learning-based congestion control  algorithms\cite{winstein2013tcp, sivaraman2014experimental}.}, when other research communities have had reproducibility embedded in their peer review processes. Reproducibility with real or emulated networks is not possible. Finally, we posit that the development of a \textit{training platform} would be broadly beneficial for the research community; we envisage a platform such as \cite{1606.01540} where researchers can formulate problems and respective RL models, train agents, and share learned policies along with all selected parameters and hyper-parameters, without having to deal with implementation details (and inevitably bugs) related to embedding learning in their simulation models.

In the past years, we have a seen a flurry of research on RL-based network technologies, e.g., in the areas of caching, routing, and congestion control, where training and evaluation is done through simulations. Although one would have expected to see the rise of simulation platforms that enable the development of said technologies, most of the literature is based on bespoke approaches; either written from scratch (e.g., as in \cite{8879573,naparstek2018deep}), or, when using an existing framework (e.g., as in \cite{10.1145/3424978.3425004,9274515}), being applicable only to the specific problem and, in most cases, not made publicly available. Lacking such publicly available and widely applicable simulation platforms is counter-productive for the research community. Bespoke approaches can be inefficient and buggy, where a widely adopted platform with standardised interfaces for model developers to interact with RL frameworks could enable robust, efficient and reproducible training and evaluation, at scale. In turn, this would enable the community to focus on the research problems and not on reinventing the wheel in combining network simulations with RL training. Notably, \textit{ns3-gym} \cite{gawlowicz2019ns} is a ns-3-based platform shares similar objectives to the work presented in this paper. Although \textit{ns3-gym} is a step to the right direction and complements our work, we have identified a number of limitations; a substantial execution overhead due to the need for inter-process communication (using \textit{ZeroMQ}) to communicate RL observations, rewards and actions between the learning framework and network simulator; fundamental limitations at the design level due to its lack of support for variable-length, dynamically adjustable steps, and multi-agent training/evaluation where agents can step independently to each other, both of which which severely constrain its applicability to real-world RL problems.

In this paper, we introduce \textit{RayNet}, a scalable and flexible platform for developing RL-based network protocols. RayNet brings together two state of the art frameworks, namely OMNeT++ \cite{varga2008overview} and Ray/RLlib \cite{moritz2018ray}, in an elegant and resource-efficient way. OMNeT++ is a state-of-the-art packet level, discrete event network simulator that is widely used by the networking research community and supports fully reproducible simulations. Ray is a general-purpose and universal distributed compute framework designed to perform any compute-intensive job (written in Python) with flexibility, including distributed training, hyper-parameter tuning, deep reinforcement learning (DRL), and production model serving. RLlib \cite{liang2018rllib} in particular is an open-source RL toolkit that employs fine-grained nested parallelism to achieve state-of-the-art performance across a wide variety of RL workloads and provides scalable abstractions for assembling new RL algorithms with minimal programming overhead. RayNet is embedded deeply within OMNeT++ by operating directly within the simulator's event loop to control the simulation when needed; i.e., to collect experience - observations and reward - and update agent actions within a simulated network environment. At the same time, RayNet operates as a Ray trainer, enabling users to run distributed multi-node training on simulated networks through a set of Python bindings. Said integration is very efficient and the only (minimal) overhead is induced by the Python bindings that allow Ray to control a network simulation. RayNet promotes code reusability by abstracting away all implementation details related to communicating RL-related information in and out of a simulation, and offering a simple API to developers, whereby they can embed RL into their models for training and evaluation. We prototype an RL-driven congestion control approach as a case study to demonstrate how RayNet facilitates the design, engineering, and evaluation of RL solutions for complex networking problems, at scale. Through experimentation, we show how RayNet enables optimisation and analysis of our RL-driven congestion control protocol, decoupling the learning logic configuration from the networking environment set-up, and providing a multi-agent simulation platform. RayNet is, to the best of our knowledge, the first platform to integrate Ray/RLlib and OMNeT++ end-to-end, providing a scalable platform for training and evaluating RL-based network protocols with minimal overhead. We compare RayNet's performance with that of \textit{ns3-gym} to demonstrate its superiority. RayNet is available as an open source project at \textit{https://github.com/giacomoni/raynet}.
 
\section{Background}
\label{background}
In this section we briefly present work related to RayNet. First, we provide an overview of RL and congestion control. Then, we discuss OMNeT++, focussing on its discrete event nature and programming interface for controlling network simulations, and, subsequently, describe key characteristics of Ray and RLlib. Finally, we present literature on RL-based research where computer network simulators have been employed, highlighting the scarcity of comprehensive, open-source platforms, such as the one we propose in this paper.

\subsection{Reinforcement Learning}
\label{rl}


RL is the process of learning how to maximise a numerical reward signal by mapping states to actions. The agent is initialised with a random decision-making strategy, thus it must experiment to determine which actions yield the greatest reward. Actions may affect not only the immediate reward but also some or all of subsequent rewards.
The problem of RL can be formalised as the optimal control of partially observable Markov decision processes (POMDP), a framing of the problem of learning from interaction to achieve a goal. The agent and environment interact at each one of a sequence of discrete time steps, $t_0,t_1,t_2,t_3,...,t_n$
At each time step $t$, the agent receives a representation of the environment’s state $S_t \in \mathcal{S}$, and selects an action $A_t \in \mathcal{A}$. One time step later, partly as a consequence of its action, the agent receives a numerical reward $R_{t+1}$, and finds itself in a new state $S_{t+1}$. The MDP and agent together give rise to a \textit{trajectory}, a sequence of states, actions and rewards $S_0, A_0, R_1, S_1, A_1, R_2, S_2, A_2, R_3, ...$. The goal for an RL agent is to find the policy, a mapping between state and actions $\pi: \mathcal{S} \rightarrow \mathcal{A}$, that maximises the discounted cumulative reward:

\begin{equation}
    \sum_{k=0}^{\infty} \gamma^k R_{t+k+1}
\end{equation}

where $\gamma \in [0,1]$ is the discount rate. DRL combines RL methods with non-linear function approximation, i.e., deep neural networks, to cope with target tasks in which encountered states may never have been seen before. In order to make rational judgments in such situations, it is required to generalise from prior experiences with distinct conditions that are comparable to the current ones in some way. 
A more detailed discussion of RL can be found in \cite{sutton2018reinforcement}.

RL algorithms require to sample trajectories of experience in order to improve the current policy. On-policy algorithms learn from experience sampled by the policy itself, whereas off-policy algorithms use (or re-use) experience collected by other policies to improve the currently optimised policy. 
Collecting trajectories of experience can be non-trivial for certain tasks. First, RL performs a trial-and-error search, and this involves the trade-off between exploration and exploitation; to gain a large cumulative reward, an agent favours actions that have been previously tried and found rewarding. However, to discover such actions, it must try actions that have not been previously chosen. In certain applications (e.g., autonomous driving), the risk associated with the exploration can be high and the agent has to be trained in a safe environment before it is deployed in the real one. Second, some tasks may span over large timescales and collecting samples of experience may be time-consuming. It is therefore very common that simulations are used to provide an efficient and safe environment for training RL agents. Agents can be subsequently be deployed `on the field' using the learned policy, which may or may not be updating based on experience collected by the deployed agent.

\subsection{Congestion Control}
\label{congestion control}

Multiple users accessing a network must share available resources - bandwidth and buffers. Network congestion is a network state characterised by increased network delay and packet loss rate, as a result of traffic going through one or more bottleneck links where the required bandwidth exceeds the available one. Network congestion results in severe degradation of users' quality of experience and must therefore be controlled. Congestion control involves end-hosts, and potentially in-network devices, and aims to maximise resource utilisation while fairly allocating resources among all users. This is commonly done on an end-to-end basis by regulating senders' transmission rate.

The decentralised nature of computer networks, coupled with the heterogeneity of application requirements and network architectures, makes congestion control an inherently complex problem. Networks and traffic patterns evolve, and new network architectures constantly emerge. This requires frequent rethinking of congestion control algorithms or the introduction of new ones. Multiple iterations of congestion control have been deployed in the wild, since the original algorithm was introduced in the Transmission Control Protocol (TCP); e.g., TCP Cubic \cite{ha2008cubic}, TCP Compound \cite{song2006compound} and MultiPath TCP \cite{paasch2014multipath}. Numerous algorithms have been proposed for data centre (\cite{alizadeh2010data, mittal2015timely, wu2010ictcp}), wireless (\cite{mascolo2001tcp, kliazovich2006tcp, shimonishi2005improving}) and satellite (\cite{akyildiz2001tcp, taleb2006refwa}) networks all of which present wildly different characteristics and challenges. Deployed congestion control protocols employ a sliding window of data packets to control the number of packets in transit and modify the size of the window in response to congestion events (e.g., inferred packet loss). Historically, there have been two categories of congestion control algorithms: loss-based and delay-based. Loss-based algorithms, such as TCP Cubic \cite{ha2008cubic} perform congestion control based on inferred packet loss; the window size increases until packet loss is inferred, at which point, the window size is decreased. Delay-based algorithms, such as TCP Vegas \cite{brakmo1994tcp}, consider an increase in packet latency as a sign of congestion and adjust the window to proactively decrease latency.

Recently, a new learning-based congestion control paradigm has gained traction, with the key argument being that congestion signals and control actions are too complex for humans to interpret and that machine-generated algorithms can provide superior policies compared to human-derived ones. An objective function then guides the learning of the control strategy. Early work in this thread included off-line optimisation of a fixed rule table \cite{winstein2013tcp, sivaraman2014experimental} and online gradient ascent optimisation \cite{dong2015pcc, dong2018pcc}, with later work adopting sequential decision-making optimisation via RL algorithms \cite{jay2019deep, lan2019deep, abbasloo2020classic}. For brevity, and in order to avoid repetition in describing RL-based congestion control approaches, we refer the reader to Section \ref{cc_with_raynet}, where we discuss in detail a representative formulation of the RL-based congestion control problem as part of the use case implemented in RayNet.

\subsection{OMNeT++ Simulator}
\label{OMNeT++}

OMNeT++ is an extensible, component-based C++ simulation framework for developing network simulators. The term `network' is used in a broad sense, encompassing wired and wireless communication networks, on-chip networks, queuing networks, etc. A simulation model consists of one or more components that encapsulate network functionality (e.g., the TCP/IP protocol stack, communication channels) and interact with each other through gates. OMNeT++ is a discrete-event simulator, which means that time progresses through scheduling events in the simulated future. The event queue (called future event set in OMNeT++) is a key data structure; the simulator executes events in chronological order, until no more events exist in the queue, or some other terminal state is reached; e.g., a predetermined maximum duration is reached or a model-specific end-state predicate is true and the simulation is ended programmatically.


An OMNeT++ simulation runs as a single-threaded process\footnote{OMNeT++ supports parallel simulations but, in this paper, we focus only on single-threaded ones. In Section \ref{conclusion} we discuss future work related to RayNet and parallel simulations.}; 
at a high level, the life cycle of an OMNeT++ simulation is as follows: the  network model is first imported and initialised, and all simulation components are declared and interconnected in a collection of descriptive files (called NED files) that constitute the model. The NED files specify the simulation model's structure, such as the number of nodes in the network, the links between nodes, and the protocols stacks supported at each node. Configuration (.ini) files set model components to work in a certain way, including the kind of application running on each node, the type of traffic, the physical link attributes. Each component is then dynamically linked to the simulation kernel. During initialisation of the network model one or more events may be instantiated and inserted in the event queue. The simulator program processes each one of the events in the queue in a chronological order, and this may involve sending a message through a network link, or executing a timeout event. Processing an event may generate new events; e.g., when a timer is reset and rescheduled for the simulated future after its expiration. OMNeT++ exports an Application Programming Interface (API) through which simulations can be executed and controlled programmatically. Through this API, a programmer can iterate over events in the queue and process them individually; this feature is key in RayNet's integration with the Ray and RLlib as discussed in Section \ref{raynet_implementation}.

\subsection{Ray and RLlib}
\label{ray_rllib}

Ray \cite{moritz2018ray} is a framework for general-purpose cluster computing that, amongst others, supports simulation, training, and servicing for RL applications. RLlib \cite{liang2018rllib} is an open-source library that provides scalable software primitives for RL and enables a broad range of algorithms to be implemented with high performance, scalability, and substantial code reuse. RLlib supports a variety of environment interfaces for training agents.\footnote{See https://docs.ray.io/en/latest/rllib/rllib-env.html for an excellent discussion of supported RLlib environments.} \textit{OpenAI Gym} is the primary interface for single-agent training environments. When an episode begins, the initial observation is returned to the agent and the environment is reset to its initial state. The agent interacts with the environment by providing the action, and it receives a reward for the action performed and the subsequent observation. This interaction takes place in \textit{steps}, until an episode termination condition is met, either because the environment has achieved a terminal state or because the environment's maximum number of steps has been reached. In a multi-agent environment, numerous agents may act simultaneously, sequentially, or in a combination of the two. The \textit{MutiAgentEnv} interface enables mapping of trajectories to individual agents and the assignment of distinct policies to distinct agents. An agent can be mapped to a single policy, while a policy can be mapped to multiple agents. 

\subsection{RL-based Research and Simulations}
\label{rlresearch_simulations}

As mentioned in the introduction, there has been a flurry of research in RL-based network technologies and many of those involve training and evaluation in simulated network environments. Here, we discuss the gap that RayNet is aspiring to fill, by providing a simulation platform with standardised interfaces so that researchers can focus on the problem at hand, and not on reinventing the wheel with bespoke approaches that are only applicable to a specific research problem and network environment, as this can be inefficient and error-prone. We have explored a large body of the literature on RL-based approaches for congestion control, routing, caching, spectrum access and signal processing systems and concisely categorise adopted simulation approaches for learning agent policies below.\footnote{For brevity, we do not include all relevant papers but cite representative ones from each category and research area.}

Given that most, if not all, learning and data manipulation frameworks are either implemented or expose APIs in Python, a large body of research is based on bespoke \textit{Python-based simulators}. Examples of such RL-based research include work on content caching in vehicular networks \cite{8879573} and the Internet of Things \cite{9238811, 9194445, 8542696}, routing \cite{8703471} and intelligent network coding \cite{9488770}. Such simulators narrowly focus on specific problems, therefore they would be of little use to other networking sub-communities even if they were made publicly available (which is the exception than the norm). Another common example of bespoke simulators are the ones based on MATLAB. Examples include spectrum access \cite{naparstek2018deep}, content caching \cite{8629363, 8964499}, routing in vehicular networks \cite{9141401} and signal processing \cite{5986747}. Again, these appear to be very narrowly focussed and without offering any reusability across problem spaces and network environments.

Several RL-based approaches for routing optimisation \cite{10.1145/3342280.3342317,10.1145/3424978.3425004,8502806} and caching \cite{10070376} have combined OMNeT++ with some learning framework, such as TensorFlow (e.g., \cite{10.1145/3424978.3425004,8502806}). However, in none of these approaches, OMNeT++ is integrated with the learning framework as in RayNet. Although source code is not available for any of those papers, it is evident that OMNeT++ is used separately to the learning framework to generate observations and rewards that are then obtained externally (presumably from a stored log output by OMNeT++), by the learning framework. For example, in \cite{10.1145/3424978.3425004}, the authors state ``given routing scheme and the traffic matrix, OMNeT++ simulation environment can obtain parameters of network performance (e.g. throughout, delay) as rewards for DDPG routing optimization mechanism''. In \cite{8502806}, the authors state ``given the traffic matrix and routing solution, we can obtain network performance parameters, such as network delay, directly from the OMNeT++''. In \cite{stampa2017deep}, a similar setup is introduced for routing optimisation but the source code is made publicly available. There, one can observe that OMNeT++ simulations run independently of the learning framework, and RL-specific data is collected from the standard output of the simulator, and subsequently fed into the learning framework. Evidently, all these approaches are bespoke to the specific problem and setup, without any integration that would enable reusability and efficiency of the framework, and therefore are not directly comparable with RayNet. Similarly to what mentioned above, a couple of approaches for RL-based routing (\cite{9274515,1420665}) employ ns-2 and ns-3 simulations; however, no source code is provided, the solution is routing-specific and there is no indication that any sort of integration between the simulator and the learning framework is attempted beyond collecting statistics at the end of a simulation run and feeding them to the learning framework.

To the best of our knowledge, the only other approach with similar aspirations to RayNet is \textit{ns3-gym} \cite{gawlowicz2019ns}. At its core, \textit{ns3-gym} implements a Python OpenAI gym environment that executes an ns-3 simulation as a separate process. It employs a socket-based inter process communication library (\textit{ZeroMQ}) to support communication between Python and ns-3 processes. This is in contrast to RayNet which integrates the simulated environment within a single process through the use of \textit{pybind11}, resulting in a much more efficient implementation. At the design level, in \textit{ns3-gym}, the environment steps at fixed simulated time intervals, which the user must set before the python process is executed; this is a particularly constraining aspect because not all problems can be formulated with fixed-length steps. Importantly, this also imposes a significant constraint on multi-agent training and evaluation. This is because with \textit{ns3-gym}, all agents need to step synchronously, but this is rarely the case in real-world problems. RayNet, on the other hand, allows for fully configurable step lengths which can be adjusted dynamically as the simulation runs, enabling support for much more complex problem setups. For example, in Section \ref{cc_with_raynet}, we show how we can formulate the congestion control problem with RayNet where RL agents, modelling network flows, can come and go, and step independently to each other. Such formulation is very important for researching RL-based congestion control, but would be impossible with \textit{ns3-gym}. Finally, \textit{ns3-gym} requires agents to start stepping at the beginning of the simulation and this can be problematic for tasks in which the agent only starts acting on the environment after a specific event occurs. This is the case in congestion control, where the RL agent starts stepping after a slow start phase (e.g., as \cite{abbasloo2020classic}), which would not be possible with \textit{ns3-gym} but is trivially supported by RayNet.

\section{Design Principles}
\label{principles}


\noindent\textbf{Separation of environment from learning.} As with standard RL training platforms, the environment must be logically separated from the learning process and its execution needs to adhere to a few simple operations (and, programmatically, to a respective API). This ensures that one can change learning algorithms and hyper-parameters without having to change anything in the environment where the agent(s) operate(s) in. Conversely, it is crucial that the environment can change without requiring any changes to the learning infrastructure, in order to support learning in different contexts and scenarios. A typical example of such a separation is the \textit{OpenAI Gym} abstraction that is widely adopted in numerous RL setups. Such a separation is important in the context of computer networks, where a single agent may have to be trained in diverse network setups, with respect to the physical topology, number of end-hosts, and traffic workloads. The step size definition may be different depending on the nature of the problem. Typically, games, such as Chess or Go, have a discrete, turn-based structure in which each step corresponds to a player's turn. Other tasks, including Atari Games and Robot Control, necessitate the discretisation of time. Agents act at predefined time intervals that signal an RL step, the length of which depends on the particular task. It is therefore important that fine-grained control of the step size is supported, where agents can step independently to each other, in groups or individually. RayNet allows for a step-based approach (built on top of the \textit{OpenAI Gym} environment) by having Ray directly controlling the event execution loop for each simulation, as discussed in Section \ref{event_looping}.

\noindent\textbf{Support for multi-agent environments.} Learning policies for core network functions, such as congestion control  and routing, requires operating multiple agents within a single environment. For example, as part of the learning process, one could have multiple TCP flows (i.e., congestion control  agents that use the same policy) competing for bandwidth on a network link. Similarly, multiple routers in a network may be acting on flows independently, following the same or different policies. It is therefore crucial to support environment execution (for training and/or evaluation purposes) with multiple agents that may or may not learn/employ the same policy. RayNet supports this by integrating Ray/RLlib's multi-agent interface with a bespoke signalling system for disseminating and collecting actions, rewards, and observations that we developed using OMNeT++'s API, as detailed in Section \ref{raynet_environment}.

\noindent\textbf{Reproducibility.} Reproducibility enables researchers to replicate published results, identify errors or limitations and propose ways forward. RL is generally hard to reproduce due to the algorithms' intrinsic variance, the environments' stochasticity, and the potentially large number of hyper-parameters that can go unreported. RayNet aims to minimise factors that can lead to non-reproducible results by employing OMNeT++ as the underlying environment for collecting experience to optimise agent policies. OMNeT++ simulations are deterministic by design, allowing for fully reproducible results; OMNeT++'s sophisticated pseudo-random number generator framework allows for controlling randomness and enabling truly independent runs of the same simulation (i.e., a RayNet environment). On top of this, Ray and RLlib, support state-of-the-art reporting of hyper-parameters, limiting non-determinism only to algorithms' intrinsic variance.

\noindent\textbf{Efficiency and scalability.} RL requires the collection of a very large amount of experience through which agents learn how to best interact with their environment. Experience is then collected into a replay buffer and learning is done by drawing an experience batch out of this buffer. Three issues are crucial in this process; (1) the environment must be quick in transforming actions to rewards, and some partially observable state; (2) the learning itself must be done efficiently; (3) and all available computational resources must be used as efficiently as possible; by enabling parallel instantiation and execution of as many environments as possible; and by minimising the overhead in the interaction between the learning and environment execution components of the RL setup. RayNet adheres to this principle, by using lightweight Python bindings (see Section \ref{overview}) to integrate Ray/RLlib with OMNeT++ in a programmatic fashion. Ray allows for running multiple environments in parallel, and OMNeT++ itself runs each environment (i.e., a network simulation) efficiently as a single-threaded process which enables parallel learning at scale.

\section{RayNet Architecture}
\label{raynet_implementation}

In this section, we discuss RayNet in detail. We first provide a high-level overview of its architecture abstracting away implementation details and focussing on the interfaces exported by the environment to model developers and, separately, to the trainer. Next, we describe how the RLlib environment is integrated with OMNeT++'s event loop and simulation models. We then focus on the bespoke signal system we developed within OMNeT++ so that Ray/RLlib can efficiently communicate actions to agents and collect observations and rewards from them at user-defined learning steps. Finally, we discuss how we can deploy trained RL agents into simulated and real-world networks.

\begin{figure}
\includegraphics[scale=0.5]{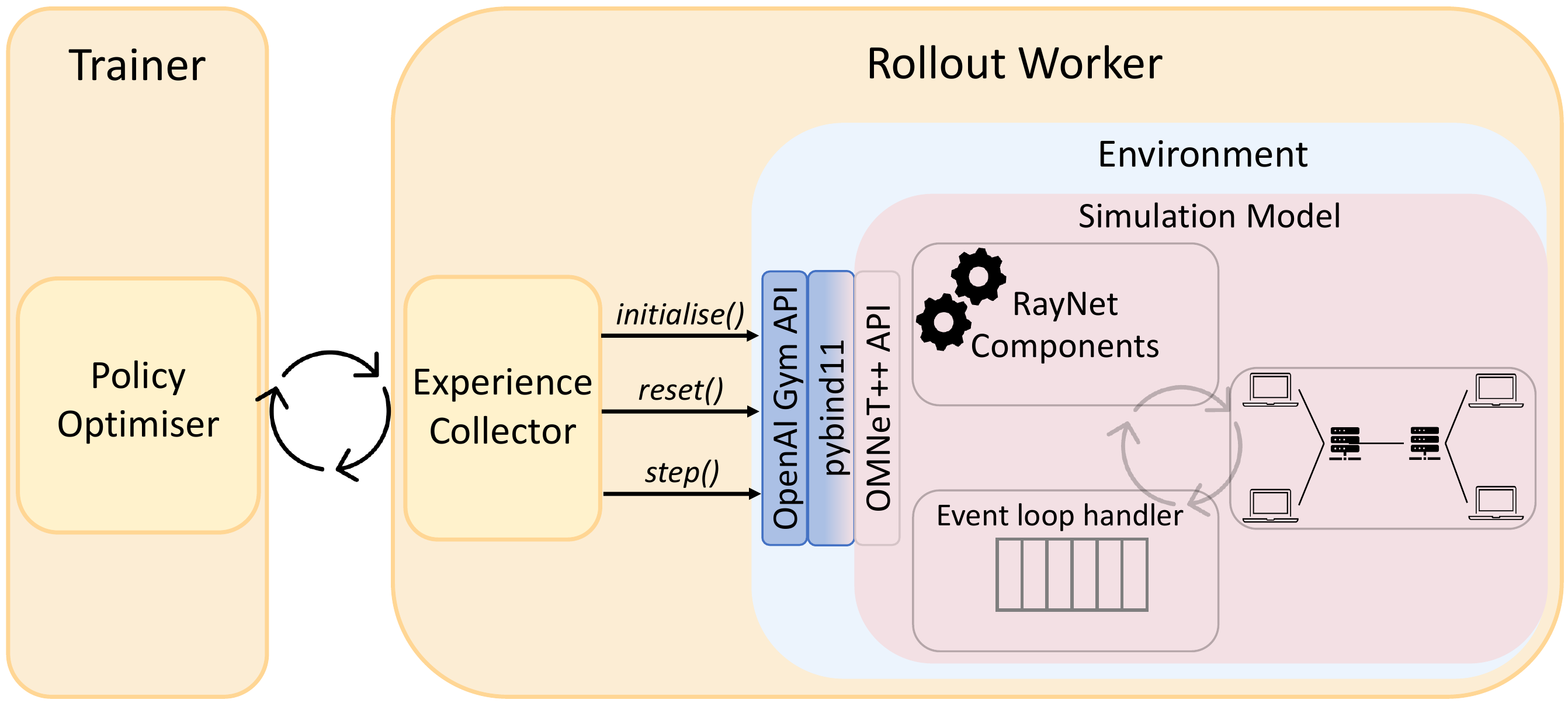}
\centering
\caption{An overview of a RayNet components. The trainer and rollout worker are Ray processes. The worker initialises and controls a Python class that implements the OpenAI Gym interface and serves as the RL environment. OMNeT++ core libraries, the event loop handler, simulation models and RayNet components are programmatically embedded into the environment through the \textit{pybind11} API.}
\label{core_modules}
\end{figure}

\subsection{Overview}
\label{overview}

Figure \ref{core_modules} illustrates a high-level view of RayNet, with an instance of the trainer (Ray/RLlib), shown on the left, and a rollout worker that encompasses the environment, shown on the right; for clarity, only a single worker is shown to interact with Ray's trainer. The trainer is the process running the policy optimiser and underlying RL algorithm, and delegates policy evaluation to one or more rollout workers, which run in parallel to speed up experience collection during training. Each rollout worker is assigned one (as in Figure \ref{core_modules}) or more RayNet environments and interacts with them by calling the methods exported by the OpenAI Gym interface, namely \textit{initialise()}, \textit{step()} and \textit{reset()}, as shown in Figure \ref{core_modules}. The environment itself is a simulated network, consisting of OMNeT++ core libraries, the event loop handler, simulation models (including the simulated network and used protocol stack) and RayNet components.

\noindent\textbf{initialise():} this method creates and initialises the environment. Objects of all core simulation classes along with classes that comprise the network are instantiated. The event loop handler is instantiated and, as part of the network initialisation, one or more events are scheduled (e.g., an application sends a packet down to the data transport layer or a wireless node broadcasts a link layer frame). Note that the simulation has not started at this point; an episode is started only after the \textit{reset()} method below is called.
    
\noindent\textbf{reset():} at any point in an episode, the worker can call this method to restore the environment to a random or starting state so that a new episode can begin. The function returns the starting observation for this new episode.
    
\noindent\textbf{step(action):} progression within an episode is done through a series of steps, cycles of state-action-reward sequences, as discussed in Section \ref{rl}. Within RayNet, the worker goes through this series of steps by repeatedly calling this method, each time providing the action to be performed on the environment; the environment, in turn, transitions to a subsequent RL state. The function returns an observation of the newly attained state, a reward value for the performed action, and a Boolean flag indicating whether the new state is final (i.e., the end of an episode) or not. Multiple RL agents can step in the environment simultaneously in which case all the aforementioned scalar values become vectors of values, with each element in the vector (e.g., the reward vector) being associated to a specific agent operating in the environment. The definition of the step, action, reward and observation, along with how the environment transitions from one state to another are problem-specific and the responsibility of the simulation model developer. The RL algorithm that runs within the trainer (see Figure \ref{core_modules}) feeds the reward and observation into its internal model, without having to know how the environment transitions; all environment-specific knowledge is embedded in the simulation model.

Within the simulation environment, access to RL functionality is provided to simulation models through the interface shown below. The developer need not worry about internal mechanics of the integration between the trainer and simulator, in terms of communicating observations, rewards and actions to and from RL agents in the environment. Such details are hidden from the developer, and implemented within RayNet, as discussed in Sections \ref{event_looping} and \ref{raynet_environment}.

\noindent\textbf{registerAgent():} each RL agent in the simulated network has to register itself with RayNet components that handle all internal communication and stepping functionality (see Sections \ref{event_looping} and \ref{raynet_environment}). At this point, the environment will not step the newly joined agent until a step length is set for it, using the method discussed below. An agent is therefore not required to start stepping immediately after joining the simulated network.\footnote{As discussed in Section \ref{cc_with_raynet}, this is important in congestion control where a TCP sender must first go through a slow start phase before the agent starts stepping.}

\noindent\textbf{setNextStep():} each RL agent is responsible for setting the duration of its own RL step; this duration may be fixed throughout an episode or updated by the agent in a problem-specific fashion; for example, a congestion control agent may be setting its next step to be a multiple of some estimated network latency, where the estimation itself changes throughout the episode. Behind the scenes, RayNet schedules special \textsc{step} events that allow the simulated environment to step in sync with the trainer, as discussed in Section \ref{event_looping}. Note that different RL agents may be stepping at different times, and can set their own step duration independently to each other.

\noindent\textbf{getObs():} this is a callback method that is implemented by each RL agent in the environment. It is called by RayNet at the end of an RL step and must produce an observation of the current environment state that is returned to the trainer. Calculating the observation is problem-specific and the trainer itself is agnostic to the semantics of the observation.

\noindent\textbf{getReward():} similarly to the \textit{getObs()} callback, this callback is called by RayNet to signify to the RL agent that the reward for the last step must be calculated and returned to the trainer. Its implementation is problem-specific.

\noindent\textbf{getDone():} this callback must be implemented by an RL agent and should return \textsc{true} if the current episode has come to an end with the completion of this step, otherwise the method should return \textsc{false}.

\noindent\textbf{setAction():} RayNet calls this callback method at the beginning of a step so that the RL agent can update its action, as dictated by the RL policy maintained by the trainer. As with the callbacks above, the semantics of the action are problem-specific and only known to the RL agent.

As discussed in Section \ref{OMNeT++}, OMNeT++ exports an API to initialise, and run simulations, programmatically, which RayNet employs to control the life cycle of a simulation; said life cycle is mapped to the methods exported by the \textit{OpenAI Gym} interface, effectively integrating Ray/RLlib with OMNeT++. In Section \ref{raynet_environment}, we discuss in detail the interactions within the OMNeT++ simulation in response to calling these methods.  The last piece of the integration is the Python bindings implemented using \textit{pybind11}\footnote{https://pybind11.readthedocs.io/}. As shown in Figure \ref{core_modules}, the bindings sit between the worker and the OMNeT++ API and implement the \textit{OpenAI Gym} methods by calling C++ methods that directly invoke the OMNeT++ API.

\begin{algorithm}
\caption{Environment Stepping}\label{embedded}
\begin{algorithmic}[1]
\Procedure{Step}{}
\State $\textit{model.setAction(action)} $
\While{ $\textit{queue.size} \neq 0$}
\State $\textit{event} \gets \textit{queue.get}$ \Comment{retrieve event from head of event queue}
\State $\textit{currentTime} \gets \textit{event.timestamp}$ \Comment{advance simulated time}
\If{\textit{event}.type = \textsc{step}}
\State $\textit{obs} \gets \textit{model.getObs}$
\State $\textit{reward} \gets \textit{model.getReward}$
\State $\textit{done} \gets \textit{model.getDone}$
\State \Return \textit{(obs, reward, done)} \Comment{return control to the worker}
\Else
\State \textbf{process(}\textit{event}\textbf{)}  \Comment{processing may alter event queue or end simulation/RL episode}
\EndIf
\EndWhile
\EndProcedure
\end{algorithmic}
\end{algorithm}

\subsection{Environment Stepping and Event Looping}
\label{event_looping}

Performing a full interaction step between the agent(s) and the environment  in RayNet requires executing one or more\footnote{In a large scale network simulation there could be thousands of simulated events in a single environment step.} simulated events. As a result of executing an event, the simulated time advances to the time at which the event was scheduled to be executed. RayNet integrates OMNeT++ into Ray workers by directly accessing its event processing loop. The pseudo-code shown in Listing \ref{embedded} illustrates this integration. Every time a worker calls the \textit{step()} method, through the \textit{pybind11} API (see Figure \ref{core_modules}), control is passed to the event loop in Listing \ref{embedded}, which iterates over each one of the events in the queue in chronological order and processes it (lines 2 - 12), until either no more events exist, a model-specific termination state is reached or a special \textsc{step} event is encountered (line 6). In the former case, the simulation is completed, and the worker cleans up the simulation outside this method. In the latter case, the end of a step is signified and the worker collects the rewards and observations from the agents whose steps in the environment have been completed (lines 7 - 9); this is done by calling the callback methods discussed in the previous section (i.e., \textit{getObs()}, \textit{getReward()} and \textit{getDone()}). We discuss how these special events are placed in the queue in Section \ref{raynet_environment}.

There are three important points related to the described integration; (1) The placement of the \textsc{step} event in the simulator's event queue effectively aligns the stepping nature of RL with the discrete event nature of the simulated environment. By requiring RL agents to explicitly define the duration of their steps by calling the \textit{setNextStep()} method (which behind the scenes results in scheduling \textsc{step} events in the simulated future), RL and simulation progress in tandum. (2) the stepping details and the internals of the environment are abstracted away from Ray/RLlib (i.e., the trainer and workers). The worker only knows to call the \textit{step()} method, which, in turn, consumes events in OMNeT++'s queue, effectively progressing the simulation (i.e., performing operations in the environment) without any need to understand the semantics of these events. (3) Crucially, the semantics of stepping are defined within the simulation model and it is the responsibility of RL agents (i.e., the developer that implements a RL agents) to place \textsc{step} events in the queue by calling the \textit{setNextStep()} method; this provides flexibility in defining the stepping semantics in a problem-specific fashion. For example, in a congestion control problem, a step may last for a fixed amount of time, or until a specific number of packet acknowledgments are received by the sender. 

\subsection{RayNet Environment}
\label{raynet_environment}

\begin{figure}
\includegraphics[scale=0.5]{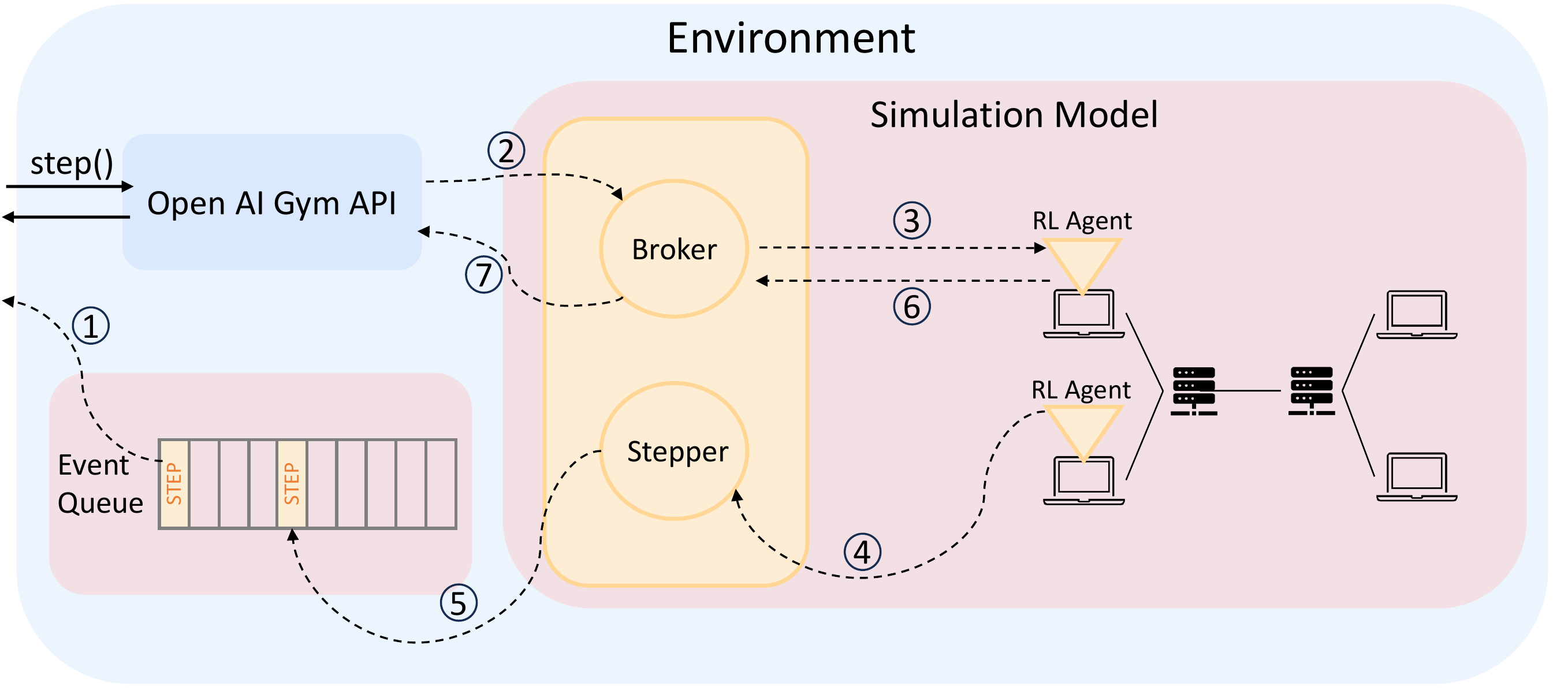}
\centering
\caption{Module interactions throughout the life cycle of a step. Components in yellow are part of RayNet's implementation. RL agents are shown as triangles and can be attached to any simulation model by implementing the API described in Section \ref{overview}}
\label{environment_internals}
\end{figure}

A RayNet environment consists of OMNeT++ simulation models, implemented as C++ modules, and RayNet-specific modules (namely \textit{RL agents}, the \textit{stepper} and \textit{broker}, as depicted in Figure \ref{environment_internals}) that facilitate environment stepping and its interaction with the Ray workers. A RayNet environment contains one or more RL agents that act based on trained policies. For example, in a congestion control setup, an agent could operate within the data transport layer, controlling the transmission rate for a specific network flow (see Section \ref{cc_with_raynet} for more details on our congestion control use case). The \textit{stepper} is responsible for coordinating with the RL agents to enable the environment to transition to a new state when the Ray worker calls the \textit{step()} method. The \textit{broker} is responsible for serialising and de-serialising action/observation/reward values (scalar or vectors) and disseminating these to agents (actions) and the Ray worker (observations/rewards), at the beginning and end of a step, respectively.

Interaction between the RayNet-specific modules is implemented using the signalling system provided by OMNeT++, which adopts a publish/subscribe communication paradigm. An OMNeT++ module can subscribe to one or more signal types by name; multiple modules can subscribe to the same signal. When a module publishes a signal of one of the types for which other modules have previously subscribed, OMNeT++ notifies these modules. A key advantage of such a paradigm is that the coupling between publisher and subscriber modules is loose; i.e., these modules do not need to know of each other to be able to communicate. This is crucial in RayNet because an environment may contain multiple agents that appear and disappear at different times during the life cycle of an environment; e.g., TCP senders for respective TCP flows. With OMNeT++'s signalling system, RL agents can communicate with the \textit{stepper} and \textit{broker} by publishing and subscribing to pre-specified signal types without referencing each other at compile time.

Upon environment initialisation, the \textit{stepper} and \textit{broker} are also initialised and, as part of this, they subscribe to specific signal types so that (1) they can coordinate with each other and (2) receive messages by RL agents. RL agents that are present when the environment is initialised, register their presence with the \textit{stepper} and \textit{broker} by publishing signals of specific types, and subscribe to specific signal types so that they can receive messages from them; RL agents that appear during the life cycle of an environment follow the same registration process with the \textit{stepper} and \textit{broker}. At this point, initialisation is complete and the \textit{initialise()} method that triggered the operations described above returns control to the Ray worker. Next, Ray workers call the \textit{reset()} method, which brings the environment to a state where the first step can be taken. This may involve processing zero to many simulated events that have been queued during initialisation and events that need to be scheduled and executed as part of the environment resetting. It is the \textit{stepper} that signals this state by inserting a \textsc{step} event at the front of the queue (i.e., scheduling an event to the present simulated time). The environment is now ready to be stepped and the \textit{reset()} method returns control to the Ray worker, along with the initial environment observation. A Ray worker subsequently steps the environment (by repeatedly calling the \textit{step()} method) until the end of the episode - the definition of the episode is problem-specific; e.g., the end of the simulation or reaching some internal milestone. The end of the episode is signified by the environment, by returning from the \textit{step()} method with the relevant Boolean flag set to true. 

Figure \ref{environment_internals} illustrates the sequence of RayNet operations that take place during an RL step. When the Ray worker calls the \textit{step()} method, the \textsc{step} event that was previously inserted in the event queue is consumed and a new RL step begins ((1) in Figure \ref{environment_internals}). The action is passed to the \textit{broker}, directly through the OMNeT++ API which allows accessing OMNeT++ modules by name ((2) in Figure \ref{environment_internals}). Depending on the number of RL agents present in the environment, this value may be a scalar value or a vector of values (in fact a pair of \{agent-id, action\} values), one for each RL agent. The \textit{broker} then broadcasts the action(s) by publishing a signal that RL agents are subscribed to ((3) in Figure \ref{environment_internals}). RayNet will then call the \textit{setAction()} callback for each one of these RL agents, at which point, they are all aware of the action they must apply in the current step. As discussed in Section \ref{event_looping}, RL agents are required to set the duration of their steps by calling the \textit{setNextStep()} method. When they do so, a signal is sent to the \textit{stepper} ((4) in Figure \ref{environment_internals}), which, in turn, schedules a \textsc{step} event for one more RL agents at the appropriate time ((5) in Figure \ref{environment_internals}). At the end of an RL step, RayNet calls the \textit{getObs()}, \textit{getReward()} and \textit{getDone()} callback methods on each RL agent for which the step ended. As a result, RL agents signal their observation and reward to the \textit{broker} ((6) in Figure \ref{environment_internals}). Finally, When the Ray worker encounters the \textsc{step} event in the event queue, it collects observations and rewards directly from the \textit{broker} through the OMNeT++ API ((7) in Figure \ref{environment_internals}). These are communicated back to the Ray trainer which places them into the RL replay buffer.

\subsection{Policy Deployment}
\label{deployment}

Deploying a learned policy within RayNet's simulated environment is straightforward; Ray/RLlib allows reloading previously trained policies, either for evaluation or resume training from some saved checkpoint. During evaluation, exploration is not necessary and can be disabled when computing actions. In fact, the reward, unless is part of the input features of the observation, is not required by the agent during the decision-making process, and it is only used during training of the policy. In Section \ref{experimental_evaluation}, we show how trained agents perform when deployed in a wide variety of environments, all of which are simulated. Deploying a policy that was learned with RayNet in a real network deployment would require (1) exporting the learned model(s) that comprise the policy which is a feature of Ray/RLlib and (2) integrating the policy with a real-world implementation of the network functionality under consideration. For example, in the congestion control  use case, one could integrate the learned policy into a user-space process that communicates with a kernel-space implementation of a modified TCP protocol using the sockets API, as it is done in \cite{abbasloo2020classic}. The modified TCP protocol collects observations and passes them to the user-space process that, in turn, feeds back to it an action calculated by taking into account the received observations.

\section{Learning Congestion Control with RayNet}
\label{cc_with_raynet}

In this section, we describe a use case of RayNet, namely congestion control with DRL, that we have developed to showcase RayNet's functionality. As discussed in Section \ref{congestion control}, a congestion control algorithm adjusts the amount of a flow's in-flight (i.e., sent but unacknowledged) data  with the aim to maximise overall network utilisation, while minimising perceived latency, and providing a fair bandwidth allocation to flows. An RL-based congestion control policy dictates an action (e.g., increasing the amount of in-flight data) in response to some signalled state from the network (e.g., an increase in the experienced packet round trip time (RTT), or receiving three duplicate acknowledgments).

\noindent\textbf{Problem formulation. }In our use case, each RL agent sits on the sender side of each data transport flow in the network. For each flow, data transmission takes place in steps; at the beginning of each step, the RL policy fixes the congestion window size to a selected value -  no updates in the congestion window size occur within a step in response to incoming acknowledgments or any other in-network signals. Depending on the path propagation delay and bandwidth, the steady-state congestion window size of a flow can range over several orders of magnitude, therefore, similarly to \cite{abbasloo2020classic}, to reduce the action space size, the policy action $\alpha$ is a multiplier applied to the current congestion window value. At the beginning of each step $t$, RL agents set the value of the $cwnd_{t}$, according to Equation \ref{cwnd_eq}.

\begin{equation}
\label{cwnd_eq}
    cwnd_{t} = 2^{\alpha} \times cwnd_{t-1}
\end{equation}

The choice\footnote{We do not discuss in detail the rationale behind selecting the specific action, observation and reward and present these here only for completeness so that we can discuss experimental results presented in Section \ref{experimental_evaluation}.} of $ \alpha $ is limited to the range [-2,2], so that the congestion window can increase by a max of four times and decrease to a max of a quarter of the current window size. RL agents infer the state of the network through observations defined as follows:
\begin{enumerate}
    \item The throughput $\mathcal{R}$ achieved in the last step over the estimated maximum bandwidth $\mathcal{R}^{max}$ of the flow.
    \item The min-max normalised smoothed round trip time $\tilde{d}$, measured in the last step, where the min and max round-trip time (RTT) values, $d^{min}$ and $d^{max}$ respectively, are estimated from the beginning of the flow.
    \item The ratio of packets lost $\mathcal{L}$ over total packets transmitted in the last step
    \item The current congestion window size.
\end{enumerate}

At each step, the reward $r$ assigned to the agent depends on the throughput $\mathcal{R}$, round trip time $d$ and loss rate $\mathcal{L}$ measured in the last step as follows:

\begin{equation}
  r=\begin{cases}
    \frac{\mathcal{R}}{\mathcal{R}^{max}} - L, & \text{if $\frac{\mathcal{R}}{\mathcal{R}^{max}} - L < 1 \land d = d^{min}$}.\\
     (\frac{\mathcal{R}}{\mathcal{R}^{max}} - L) \cdot \frac{d^{min}}{d} \cdot (1 - \tilde{d}), & \text{otherwise}.
  \end{cases}
\end{equation}

where $\tilde{d} = \frac{d - d^{min}}{d^{max} - d^{min}}$. \newline

\begin{figure}
\includegraphics[scale=0.7]{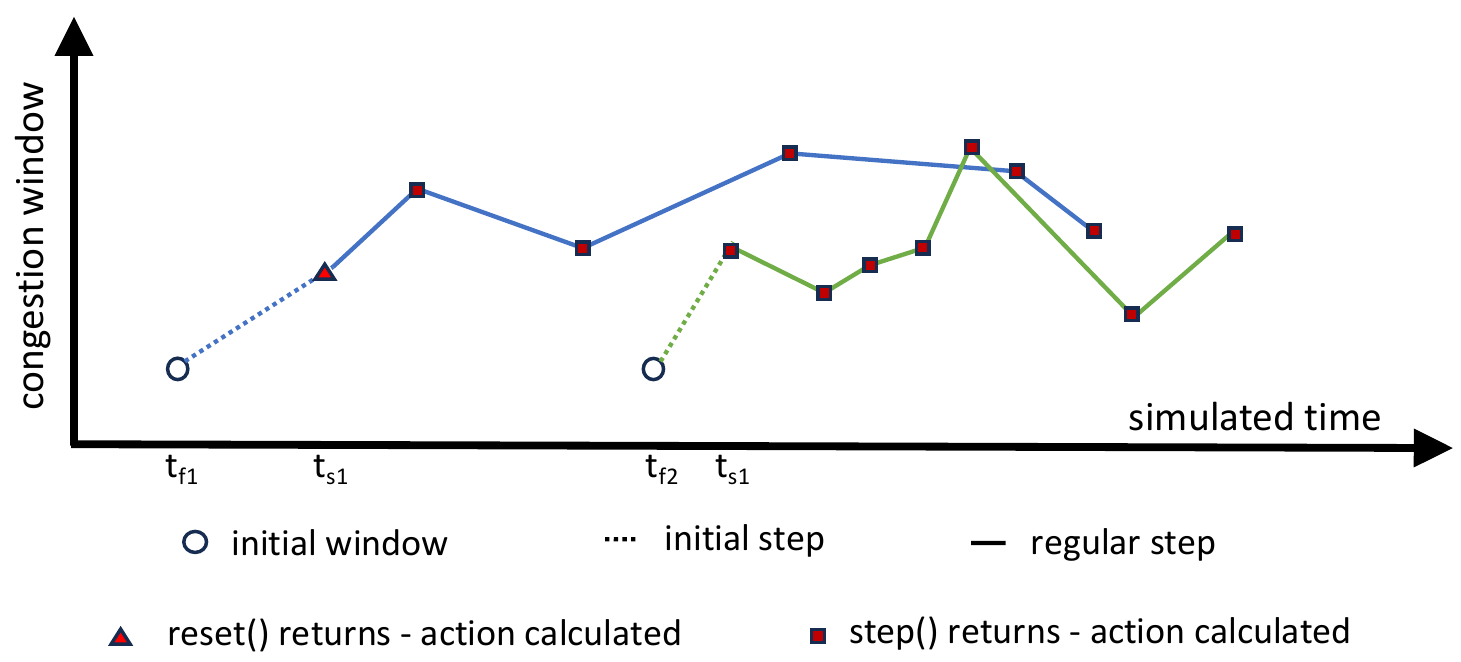}
\centering
\caption{Congestion window evolution in a simulated episode with two flows (agents).}
\label{ccusecase}
\end{figure}

\noindent\textbf{An example episode. }The life cycle of an example episode with two flows (and therefore two RL agents) is illustrated in Figure \ref{ccusecase}. After the environment initialisation, Flow 1 (shown in blue) is scheduled to start first at time $t_{f1}$ and Flow 2 (shown in green) is scheduled to start at time $t_{f2}$. Upon calling the \textit{reset()} method, the simulation loop starts executing events until the first flow starts. We employ a \textit{slow start} phase (as TCP does), during which the congestion window is initially set to a small fixed value (as in \cite{ha2008cubic}), and its size is increased exponentially in every RTT, until either a threshold is reached or packet loss occurs. This enables the sender to acquire good estimates of the current RTT, the minimum observed RTT and the maximum observed throughput; the former is used to set the duration of the first step, and the rest are used in calculating rewards. When the slow start phase ends, the flow publishes a signal to the \textit{stepper} to register its presence and calls \textit{setNextStep()} to declare the duration of its initial step. At time $t_{s1}$ the RL agent calculates its environment observation for the initial step and publishes it to the \textit{broker}, returning from the \textit{reset()} method. After inferring the next action from the received observation, the Ray worker calls the \textit{step()} method passing the new action to the environment. Each RL agent (one for each flow) in the network steps independently, and each step lasts for an amount of simulated time equal to twice the minimum RTT that the sender has observed in the last 10 seconds for this flow. Each RL agent declares its own step duration to the \textit{stepper}, by calling the \textit{setNextStep()} method; note that the step duration can vary across the lifetime of a flow, depending on the network conditions, as depicted in Figure \ref{ccusecase}. Flow 2 (shown in green) starts at time $t_{f2}$, in the middle of one of Flow 1's steps. Flow 2 also performs its initial step; at time $t_{s2}$, and before Flow 1 steps, the first observation value is returned and a new step for Flow 2 begins upon receiving a new action drawn from the RL policy. Note that RL agents of different flows step independently from the each other collecting observations and rewards that are all fed into the same training process; i.e., learning a single RL policy.

\begin{lstlisting}[caption={Programming Excerpt for RL-based Congestion Control},label={lst:excerpt}, language=C++]
class CCAlgorithm : RLInterface {        // agent inherits from RL interface (see Section 4.1)

  int cwnd = 1;                          // congestion window value (in packets)
  simtime_t d_min, d_max, d;             // minimum/maximum observed and current RTT values
  double R, R_max, L;                    // current and maximum observed throughput and loss rate
  
  void receivedAck() {                   // called by simulation model every time an ACK is received
    if (isSlowStartFinished()) {         // check if slow start phase is finished
      if(!isAgentRegistered()) {         // check if agent is registered
        registerAgent();                 // register agent with RayNet
        setNextStep(2 * d_min);          // define length of the first step
      }
      
      updateNetworkVariables();          // update d_min, d_max, d, R, R_max and L

    } else {
      performSlowStart();                // cwnd = cwnd + 1;
      updateNetworkVariables();          // update d_min, d_max, d, R, R_max and L
    }
  }
  
  // at the end of step (length defined in line 11 or 32), RayNet calls the methods below
  
  ObsType getObs() { ... }               // return observations defined in the list above
  RewardType getReward() { ... }         // calculate reward as per Eq. 3
  bool getDone() { ... }                 // TRUE if L >= threshold or all packets acknowledged

  // before the beginning of a new step, RayNet calls the method below
  
  void setAction(ActionType action) {    // called by RayNet just before a new step begins
    cwnd = (2 ^ action) * cwnd           // apply action to agent's congestion window as per Eq. 2
    setNextStep(2 * d_min);              // define length of the next step
  }
}
\end{lstlisting}

\noindent\textbf{Programming RL-based congestion control in RayNet.} Here, we briefly discuss how the RL-based congestion control use case can be implemented in RayNet. For clarity, we abstract away a lot of implementation details and focus on the core RayNet methods and their interplay with how the sender (and RL agent) handles acknowledgments. This is illustrated in the programming excerpt shown in Listing \ref{lst:excerpt}. The congestion window, which is the control variable set by the agent, is defined in line 3.  All variables that are being measured throughout the lifetime of the connection are defined in lines 4 and 5. These are used to calculate observations and rewards, as discussed above. \textit{receivedAck()} (line 7) is called by OMNeT++ every time an acknowledgment is received by the sender; this is where the core of the congestion control takes place. As discussed above, in our example, the agent is activated after the slow start phase is completed. During slow start the window is increased by one packet every time an acknowledgement is received, effectively doubling its size every RTT (line 17). Subsequently, the agent is registered with RayNet (line 10) and the length of the first step is set (line 11). Behind the scenes, RayNet steps the environment, as discussed in Sections \ref{event_looping} and \ref{raynet_environment}. Before the end of the first step, RayNet calls the three callback methods defined in lines 24, 25, 26, and feeds the returned values to the trainer. Upon the beginning of the next step, the \textit{setAction()} method in line 30 is called and the congestion window is set accordingly (line 31). The length of the step is also set (line 32). The same sequence of calls is repeated for every subsequent step, until the end of the episode, i.e., when \textit{getDone()} returns TRUE; in our example, this is when all packets have been successfully transmitted or when the loss rate reaches a predefined threshold.

\section{Experimenting with RayNet}
\label{experimental_evaluation}

In this section, we explore RayNet's capabilities and performance characteristics through experimentation with (1) the congestion control use case discussed in Section \ref{cc_with_raynet} and (2) a simple CartPole environment \cite{barto1983neuronlike} that we also developed in RayNet. Our aim is to showcase that RayNet meets the design principles set out in Section \ref{principles}.\footnote{We do not discuss the \textit{reproducibility} principle any further here; OMNeT++ simulations are deterministic therefore any reproducibility limitations only stem from RL algorithms' intrinsics.} First, we demonstrate how RayNet's design facilitates learning in the context of a complex networking task which involves the search and optimisation of RL algorithms and respective hyper-parameters, evaluation on diverse networking environments, and analysis of multi-agent performance. Second, we show evidence that RayNet's computational overhead is minimal compared to our selected baseline (Open AI Gym) when performing the same basic task (i.e., the CartPole task) in a varying amount of CPU cores. Moreover, we show that RayNet performs better compared to \textit{ns3-gym} in the same setup. Finally, we showcase RayNet running on a distributed computing environment using the congestion control task. Experiments were conducted on a Linux server with 64 CPUs and 128GB of RAM and on Amazon AWS.

\subsection{Separation of environment from learning}
\label{experimentation_separation}

\begin{figure}
\begin{minipage}[b]{0.47\textwidth}
\centering
\includegraphics[width=0.8\textwidth]{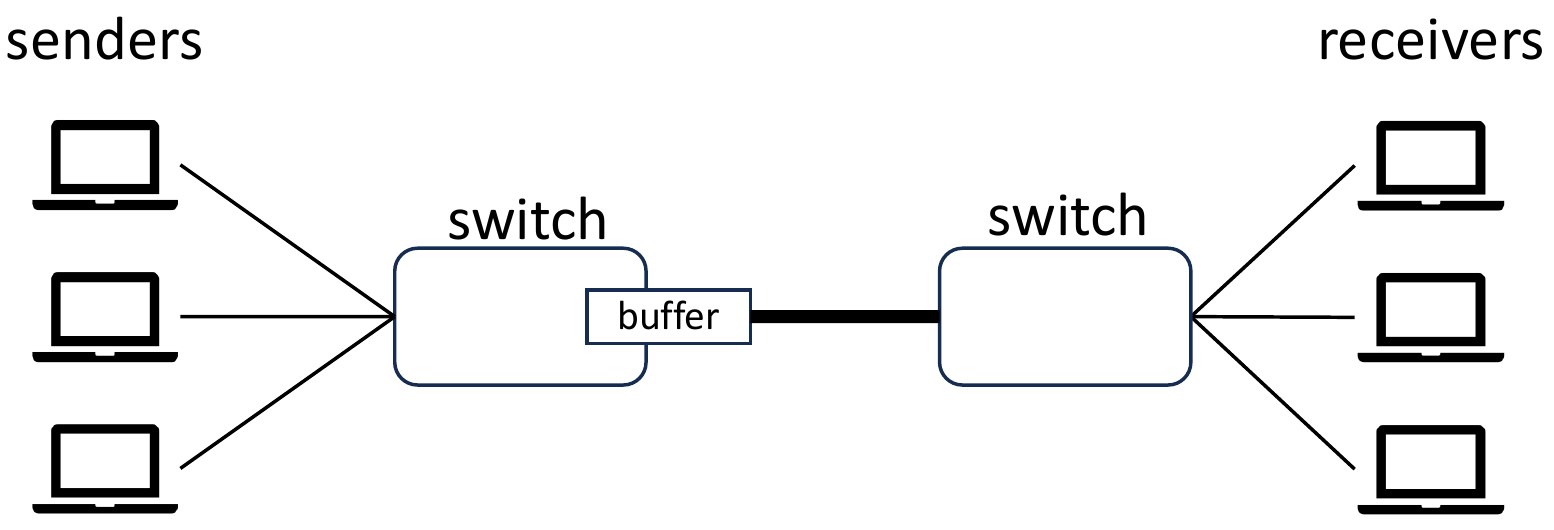}
\caption{Dumbell topology.}
\label{dumbel_topology}
\end{minipage}
\hfill
\begin{minipage}[b]{0.47\textwidth}
\centering
\begin{tabular}[t]{||c | c | c||} 
 \hline
 \textbf{Bandwidth} & \textbf{RTT} & \textbf{Buffer} \\ [0.5ex] 
 \hline\hline
 64-128Mbps & 16-64ms & 80-800 packets \\ [0.5ex] 
 \hline
\end{tabular}
\tabcaption{Network Parameters' ranges during training}
\label{tab:training_range}
\end{minipage}
\end{figure}

Here, we train models to yield efficient congestion control policies using the reward function and observations discussed in Section \ref{cc_with_raynet}. Specifically, we train a single RL agent (sender) on a Dumbell network shown in Figure \ref{dumbel_topology}. We adopt a `train and deploy' approach for our congestion control policy.

\noindent\textbf{Varying environment parameters.} We showcase how RayNet enables varying the environment completely independently of its learning components. We expose the congestion control agent to a variety of network configurations, with network parameters sampled from the ranges shown in Table \ref{tab:training_range}. We train the agent for $1$ million steps using sixteen parallel rollout workers. Each worker creates its own RL environment, uniformly sampling its parameters over each of the parameter ranges listed in Table \ref{tab:training_range}. 
Parallel experience gathering from multiple networking scenarios prevents model overfitting and avoids ``catastrophic forgetting'' of network scenarios for which experience had been gathered earlier in the training process \cite{abbasloo2020classic}. We define the parameterised networks using OMNeT++'s NED language and set the values of bottleneck bandwidth, propagation delay, and buffer size at the beginning of each episode. The agent is trained using deep deterministic policy gradient (DDPG), a model-free actor-critic algorithm based on the deterministic policy gradient that can operate over a continuous action space \cite{lillicrap2015continuous}, in conjunction with distributed prioritised experience replay \cite{horgan2018distributed}. DDPG's hyper-parameters are set to RLlib's default values

We evaluate the DDPG-trained policy on networks with parameters sampled from wider ranges than the ones used for training so that we can assess how well the policy generalises to unseen environments. 
Assessing the model's performance in regions of the state space that were not observed during training can help prevent deployment failures. We evaluate the agent by varying one of the three studied dimensions (bandwidth, propagation delay, and buffer size) within a range that includes but is broader than the respective training range (see Table \ref{tab:training_range}), while keeping the other two fixed at the mean value of the respective training range shown in Table \ref{tab:training_range}. We assess the performance of the congestion control policy by measuring three key performance metrics; \textit{normalised throughput}, the measured throughput over the theoretical maximum one, \textit{queuing delay} at the network bottleneck, and \textit{packet loss rate}.

\begin{figure}[t]
\centering
   \includegraphics[width=1\linewidth]{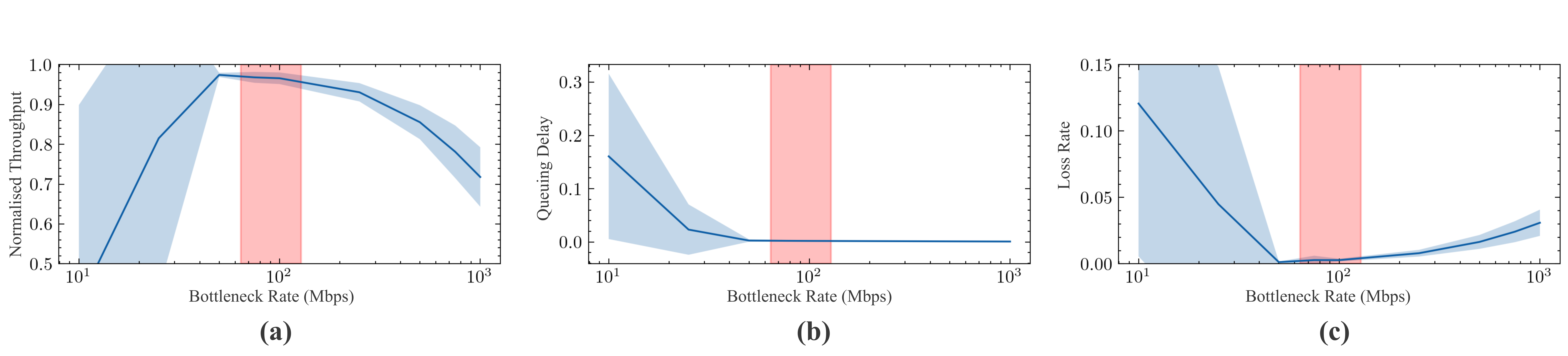}
   \caption{Normalised throughput, queuing delay and average loss rate of a single flow as the bottleneck bandwidth varies. Shaded region indicates the bottleneck bandwidth range used during training.}
   \label{fig:bandwidth} 

\centering
   \includegraphics[width=1\linewidth]{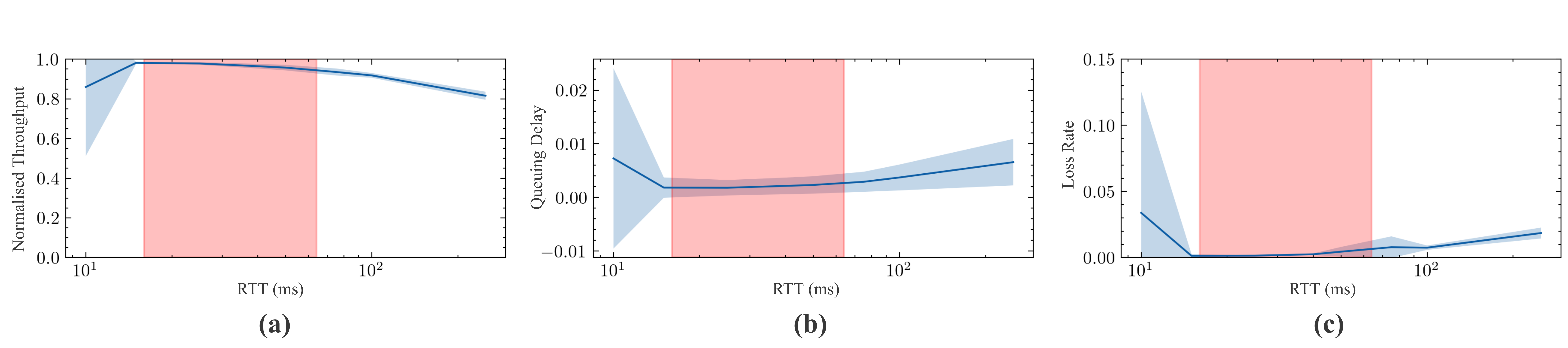}
   \caption{Normalised throughput, queuing delay and average loss rate of a single flow as the propagation delay varies. Shaded region indicates the propagation delay range used during training. }
   \label{fig:rtt}

\centering
   \includegraphics[width=1\linewidth]{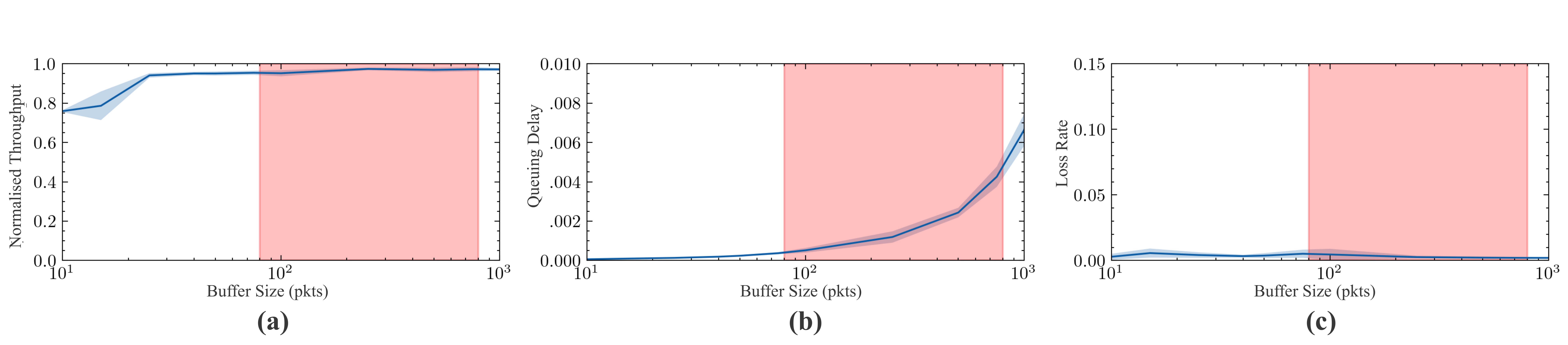}
   \caption{Normalised throughput, queuing delay and average loss rate of a single flow as the buffer size varies. Shaded region indicates the buffer size range used during training. }
   \label{fig:buffer}
\end{figure}

Figures \ref{fig:bandwidth}, \ref{fig:rtt} and \ref{fig:buffer} depict the selected performance indicators for a single flow when varying the bandwidth, propagation delay, and buffer size, respectively. Lines and blue-shaded regions denote mean value and standard deviation, respectively. On the x- and y-axis we show the varying parameter and measured performance metric, respectively. The red-shaded region depicts the training range for the varying network parameter. Although analysing these results is beyond the scope of this paper, for completeness, we briefly discuss the behaviour of the learned policy, noting that it is RayNet that allows such analysis through its clean separation of the environment from the learning components.

The bottleneck bandwidth influences throughput, queuing delay and loss rate the most, compared to propagation delay and buffer size. In fact, if the bandwidth of the network falls within the training range values, the flow achieves the highest throughput (Figure \ref{fig:bandwidth}.a), lowest queuing delay (Figure \ref{fig:bandwidth}.b) and negligible packet loss rate (Figure \ref{fig:bandwidth}.c). If the bandwidth falls outside of the training range, throughput degrades when the available bandwidth is less than the experienced values (Figure \ref{fig:bandwidth}.a), due to large congestion windows that result in overfilling the buffer, increasing queuing delays (Figure \ref{fig:bandwidth}.b) and packet loss (Figure \ref{fig:bandwidth}.c); when the available bandwidth is greater than the experienced range, we observe underutilisation of the bottleneck link, characterised by low throughput (Figure \ref{fig:bandwidth}.a) and no queuing delay (Figure \ref{fig:bandwidth}.b). Even when available bandwidth is underutilised, the flow experiences loss (Figure \ref{fig:bandwidth}.c). This is due to an increase packet loss at the slow start phase in larger bandwidth delay product (BDP) flows. When the propagation delay is raised, the resultant behaviour is comparable (Figure \ref{fig:rtt}.c). Setting the propagation delay outside of the training ranges does not affect performance as much as the variation of bandwidth does. With increased propagation delay, BDP and loss rate towards the end of slow start rise (Figure \ref{fig:rtt}.c). Queue build-ups are also more likely (Figure \ref{fig:rtt}.b). Variation in buffer size mostly increases queuing times (Figure \ref{fig:buffer}.b).

\noindent\textbf{Varying learning parameters and hyper-parameters.} Discovering and optimising RL policies often requires empirical evaluation to identify the best RL algorithm and hyper-parameters.
For instance, some tasks may inherently require a stochastic decision making policy to maximise the objective, like in the Rock-Paper-Scissors example, where a deterministic policy would inevitably lead to sub-optimal decisions. Even after fixing the RL algorithm, one needs to evaluate its efficacy within a broad range of hyper-parameters associated to the selected algorithm. For example, DDPG requires fine tuning of the exploration strategy \cite{lillicrap2015continuous}; Soft Actor-Critic (SAC) enforces exploration including the maximisation of the policy's entropy in the reward formulation, at the cost of introducing a new temperature hyper-parameter \cite{haarnoja2018soft} that trades off exploration and rewards; Proximal Policy Optimization (PPO) uses a surrogate loss function to keep the step from the old policy to the new policy within a safe range, which requires either a clipping threshold, a weighted Kullback-Leibler (KL) divergence factor or a combination of the two \cite{schulman2017proximal}. Several hyper-parameters are common to multiple RL algorithms but need to be optimised separately for each algorithm and task at hand. For example, off-policy algorithms often store experience in replay buffers, whose size must be set; the discount factor $\gamma$ is a common hyper-parameter to the majority of RL algorithms implementations; function approximators like neural networks bring their whole package of hyper-parameters, such as learning rates, loss optimizers, activation functions, and so on. RayNet supports such exploratory studies out of the box, by integrating Ray/RLlib with OMNeT++ so that the RL environment is completely separated to the learning components. To demonstrate this capability, we train RL policies in the experimental setup discussed above, using three state-of-the-art algorithms; PPO \cite{schulman2017proximal}, a policy gradient algorithm, DDPG \cite{lillicrap2015continuous}, a deterministic policy gradient algorithm with distributed prioritised experience replay, and SAC \cite{haarnoja2018soft}, a soft policy optimisation version of the actor-critic algorithm. All three algorithms are part of RLlib \cite{liang2018rllib} and all relevant configuration is done using RLlib's APIs, completely independently of the underlying environments (which are parameterised as described in the previous section).

\begin{figure}[t]
\centering
\begin{minipage}[t]{.32\textwidth}
  \centering
  \includegraphics[width=\linewidth]{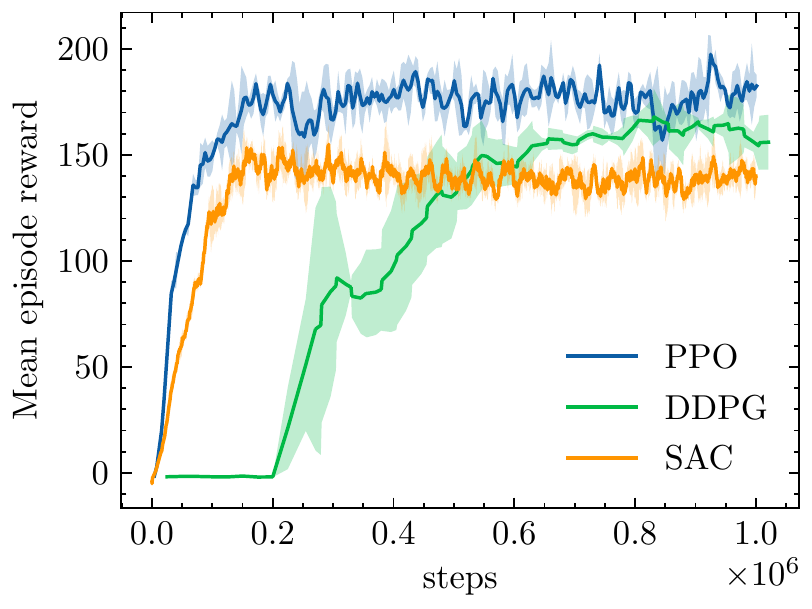}
  \captionof{figure}{Cumulative episode reward during training over the number of steps.} 
  \label{fig:episode_reward_mean}
\end{minipage}
\hfill
\begin{minipage}[t]{.32\textwidth}
  \centering
  \includegraphics[width=\linewidth]{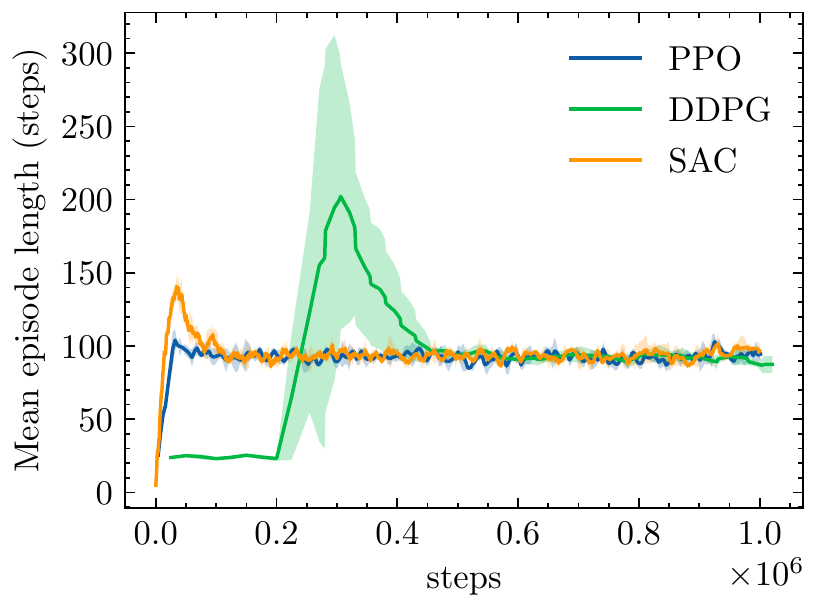}
  \captionof{figure}{Episode length during training over the number of steps. } 
  \label{fig:episode_len_mean}
\end{minipage}
\hfill
\begin{minipage}[t]{.32\textwidth}
  \centering
  \raisebox{0.3cm}{\includegraphics[width=\linewidth]{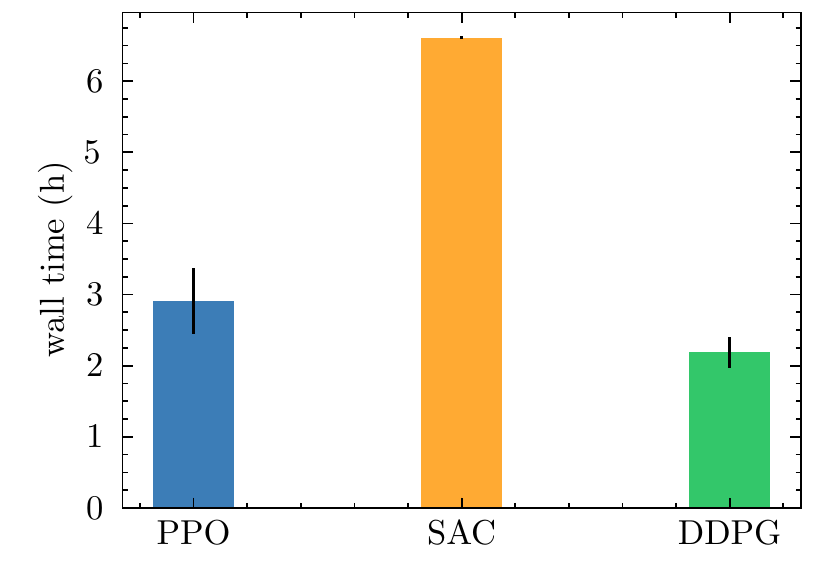}}
  \captionof{figure}{Duration of a training session of 1 million steps with PPO, SAC and DDPG.} 
  \label{fig:alg_len}
\end{minipage}
\end{figure}

Figure \ref{fig:episode_reward_mean} illustrates the average cumulative reward of episodes attained during training of the congestion control agent. Both PPO and SAC optimise a stochastic policy, that is, a distribution of actions given the state, and each policy update is constrained by selecting a safe region for the policy update (PPO) or by imposing the maximum entropy principle (SAC). In both instances, the cumulative reward is increasing monotonically, and both algorithms converged to their asymptotic optimum after the same number of training steps (around 125K). The asymptotic cumulative reward of SAC is dependent on the entropy weight factor, and the default entropy maximisation strategy led in a lower cumulative reward than PPO. DDPG, meanwhile, optimises a deterministic strategy. The policy update is not restrictive, and the policy change is contingent on the exploration strategy. In Figure \ref{fig:episode_reward_mean} we observe that DDPG is slower in learning compared to PPO and SAC, because it requires an initial warm-up time of 200K steps where experience is randomly generated. As the policy begins training, the cumulative reward reaches a local maximum before recovering and converging to a superior asymptotic maximum. This behaviour is a result of the varied duration of episodes. As per the definition of our congestion control model, an episode can end in three ways: (1) the policy generates a very high level of congestion indicated by a high amount of inferred loss; (2) the flow completes successfully; or (3) a maximum number of steps is reached (400 steps in our case). In the second scenario, the cumulative reward may be deceptive regarding the quality of the policy itself. In Figures \ref{fig:episode_reward_mean} and \ref{fig:episode_len_mean} we observe that longer episodes may result in greater cumulative reward, but for a fixed size flow, shorter episodes imply shorter flow completion times, a consequence of well behaving congestion control policies. Among the three classes of algorithms, SAC requires the longest wall clock time to complete training, twice as much compared to PPO and DDPG (Figure \ref{fig:alg_len}).

\subsection{Support for multi-agent environments}
\label{eval_multi_agent}

\begin{figure}[t]
	\centering
	\begin{minipage}[t]{.45\textwidth}
		\centering
		\includegraphics[width=.85\linewidth]{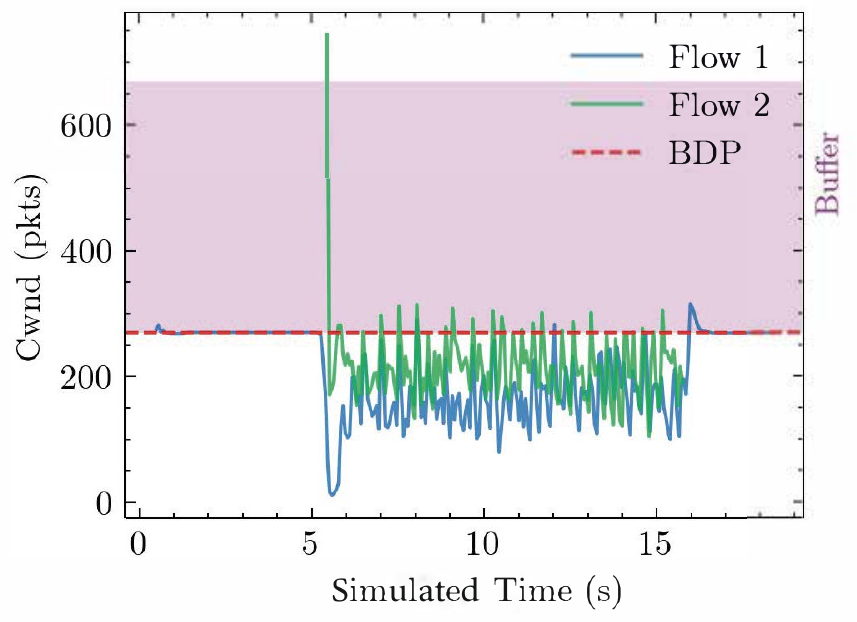}
		\caption{Evolution of the congestion window of two contending flows (agents) using the same policy.}
		\label{fig:cwnd_multi}
	\end{minipage}
	\hfill
	\begin{minipage}[t]{.45\textwidth}
		\centering
		\includegraphics[width=.85\linewidth]{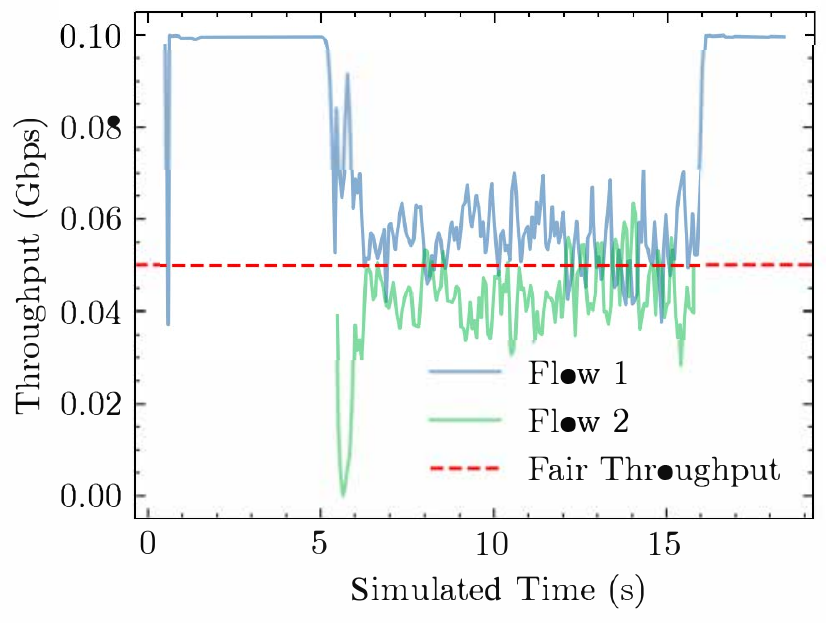}
		\caption{Throughput achieved by two contending flows (agents) using the same policy.}
		\label{fig:thr_multi}
	\end{minipage}
\end{figure}

In this section, we showcase RayNet's support for multi-agent environments by experimenting with networks where two flows contend for resources at the shared bottleneck shown in Figure \ref{dumbel_topology}. 
We run the two flows on a network with a $440$-packet buffer, a $100$Mbps bottleneck and $35$ms propagation delay. Figure \ref{fig:cwnd_multi} shows the evolution of the congestion window size for the two flows. While the first flow (blue line) is the only flow in the network, its RL agent regulates its congestion window to match the path's BDP, i.e., the optimal window size for a single flow traversing an empty network path \cite{kleinrock2018internet}. Any larger value for the congestion window and data would be buffered in the bottleneck's buffer, increasing the latency experienced by the flow. This is illustrated by the blue line coinciding with the red one in Figure \ref{fig:cwnd_multi}, for the first $5$ simulated seconds; the purple shaded region denotes congestion control values that would result in buffering. When the second flow (green line) starts, its window grows exponentially as part of the slow start phase, until loss occurs. Then, the RL agent takes over control and the congestion window is adjusted so that the cumulative value of the congestion window roughly matches the path's BDP. In Figure \ref{fig:thr_multi}, we observe that despite the lack of multiple flows experience during training, the policy achieves a relatively fair allocation of bandwidth among the two flows. When the second flow ends, the first one quickly ramps up its congestion window to acquire all available bandwidth.

\subsection{Efficiency and Scalability}
\label{experimentation_efficiency}


So far we have showcased that RayNet enables rich experimentation with RL-based network protocols, by separating the environment from the learning and supporting multiple agents when learning and evaluating RL policies. It is crucial that such features do not come at a cost of slow and resource-hungry learning. In this section, we provide evidence of RayNet's efficiency in learning RL policies taking advantage of its inherent parallelisation capability and minimal overhead when communicating environment observations, rewards and actions between Ray and OMNeT++. To do so we perform two different experiments: (1) We implement the CartPole task \textit{CartPole-v1} as an OMNeT++ model within RayNet, and within \textit{ns3-gym}, and compare the learning efficiency of these two models with that of the CartPole task implemented in the Python-based Open AI Gym. Note that our intention here is not to show any benefits of using RayNet or \textit{ns3-gym} over Open AI Gym, as the latter serves a very a different purpose. Instead, we implement the simplest possible model (i.e., the CartPole) to eliminate, as much as possible, the computational overhead from the simulation model itself, and use Open AI Gym as the performance baseline. (2) We deploy RayNet on Amazon AWS and measure its learning efficiency in a distributed computing setup; i.e., when varying the number of virtual machines used in training the underlying RL model.

\noindent\textbf{Learning with the CartPole task on a single server. }In the CartPole task, a non-actuated junction connects a pole to a frictionless track-traveling cart. The pole is positioned vertically on the cart, and the objective is to balance it by applying forces to the cart's left and right sides. 
The dynamics of the CartPole \cite{1606.01540} state transitions are implemented within a single OMNeT++ and ns-3 simulation component, for RayNet and \textit{ns3-gym}, respectively. In the RayNet model, during environment initialisation, the \textit{broker}, \textit{stepper} and CartPole modules register to specific signal types so that actions, observations, and rewards can be exchanged. When \textit{reset()} is called on the environment, the internal state of the CartPole component is set to a random value, within the defined state space. The CartPole component immediately sends the randomly generated observation to the \textit{broker}, and the \textit{stepper} inserts a \textsc{step} event into the queue. After retrieving the observation from the \textit{broker} and calculating a new action, the rollout worker invokes \textit{step()} on the environment indicating the newly calculated action to be applied by the RL agent. The \textit{broker} delivers the action to the CartPole component through a signal, triggering the state transition and reward calculation. The new observation and reward are pushed to the \textit{broker}; the \textit{stepper} inserts a new \textsc{step} event into the queue, and the \textit{step()} method completes. We have implemented the CartPole as single component within \textit{ns3-gym}, complying with its API \cite{gawlowicz2019ns}.

\begin{figure}[t]
    \centering
    \begin{minipage}[t]{.45\textwidth}
        \centering
        \includegraphics[width=0.85\linewidth]{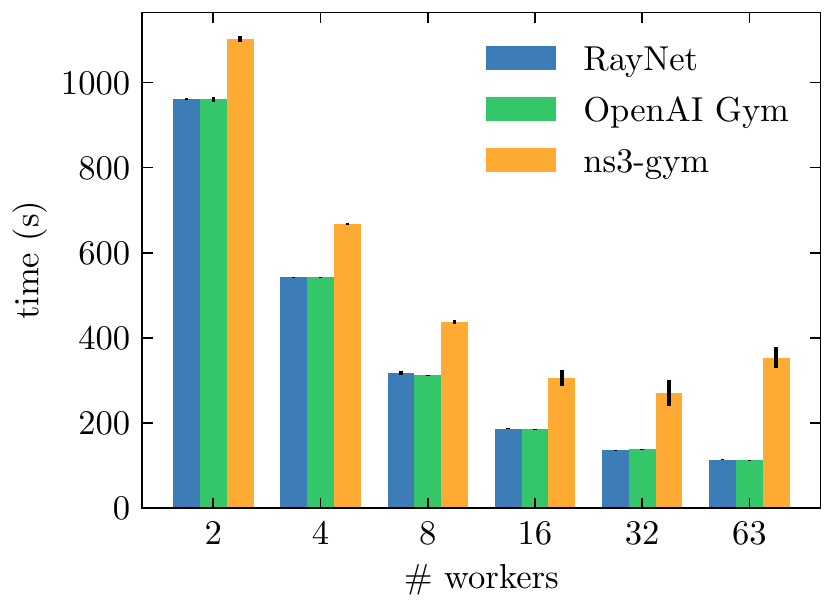}
        \captionof{figure}{Wall clock time when training a DQN policy for the CartPole task for 100K steps with a varying number of parallel workers running on a single server.}
        \label{fig:time_on_workers}
    \end{minipage}
    \hfill
    \begin{minipage}[t]{.45\textwidth}
        \centering
        \includegraphics[width=0.85\linewidth]{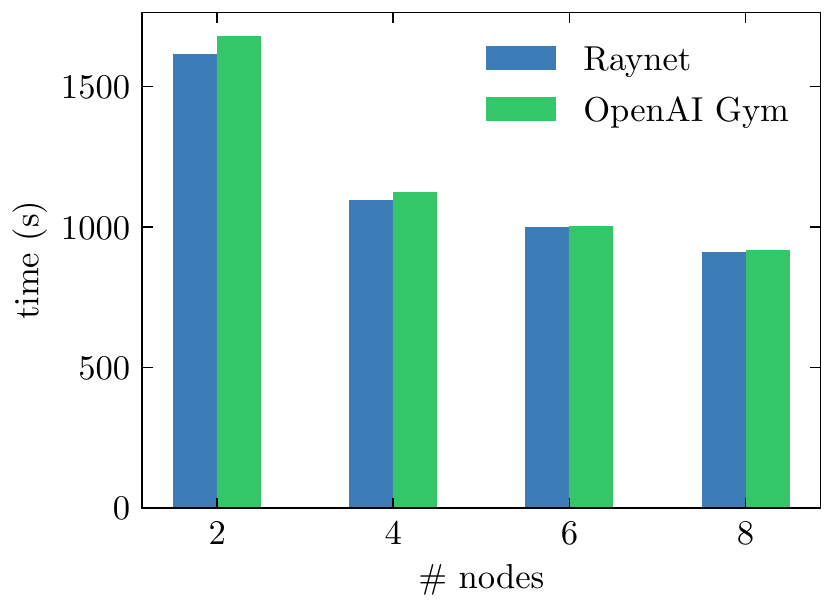}
        \captionof{figure}{Wall clock time when training a DQN policy for the CartPole task for 500k steps in a distributed Ray cluster (running on Amazon AWS) with varying number of nodes.}
        \label{fig:time_on_workers_aws}
    \end{minipage}
\end{figure}

For all three models, we train a Deep Q-Network (DQN) \cite{mnih2013playing} policy using a varying number - between 2 and 63 - of rollout workers that operate in parallel on a single server, pinned on one of the available CPU cores. The trainer itself always runs on a separate CPU core. For each <environment, number of workers> pair, we train the agent using three different seeds and each run terminates upon collecting $100$K steps. In Figure \ref{fig:time_on_workers} we observe that RayNet performs identically to the baseline Open AI Gym implementation, i.e., the overhead of introducing the integration of Ray/RLlib and OMNeT++ is negligible. This is for all numbers of workers we experimented with. Secondly, \textit{ns3-gym} consistently requires more time (more than $2\times$ as the number of workers increases) to collect the same amount of experience from the same environment. We can only attribute this to the required IPC between the ns-3 simulation and the separate Python process that is managed by Ray's rollout worker. In \textit{ns3-gym}, IPC is required for communicating actions, rewards and observations between the trainer and rollout worker and the embedded network simulator.

\noindent\textbf{Learning with the CartPole task on Amazon AWS.} We now train a policy for the CartPole task using Raynet and the Open AI Gym implementation on a distributed cluster running on Amazon AWS. The cluster is composed of nodes, each one having 8 vCPUs and 16GB of memory. We train the policy using DQN over $500$K steps, with a number of rollout workers equal to the total number of CPUs in the cluster minus one, as one CPU is reserved for the trainer process, as in the single server setup discussed above. For this experiment, it was not possible to include \textit{ns3-gym} as it does not support running on a distributed cluster. As shown in Figure \ref{fig:time_on_workers_aws}, RayNet performs equally well to Open AI Gym, accelerating learning as the number of underlying servers (and vCPUs) allocated to the task increases). This is evidence that RayNet could offer a scalable platform for training RL policies for computer network problems.

\section{Conclusion}
\label{conclusion}

In this paper, we presented RayNet, a simulation platform for training and evaluating RL-based network protocols. RayNet integrates OMNeT++, a widely used, off-the shelf discrete-event simulator, and Ray/RLlib, a distributed framework for RL at scale. The integration is achieved through the usage of Python bindings and the signalling system that is exported by OMNeT++, that allow RL agents to control decision making in the simulated environment. RayNet aspires to be a simulation platform that encompasses a diverse set of RL-based research problems and needs within the computer networks realm. It does so by abstracting away all implementation details related to communicating RL-related information in and out of a simulation, and offering a simple API to model developers by which they can embed RL into their models. Our results show how RayNet’s design facilitates learning-based protocol development; it allows separate and extensible configuration for the learning algorithm and the networking environment; it supports multi-agent simulation, with each agent stepping independently; and it introduces minimal overhead compared to existing general-purpose RL training platforms, outperforming \textit{ns3-gym} with which it shares similar objectives.

Our future work is on developing more use cases for RayNet, including RL-based routing and traffic engineering, and improving the codebase in terms of code reusability and user friendliness. RayNet is available as an open-source project for the research community to use and develop further. We are currently using RayNet in researching fair congestion control algorithms and studying existing RL-based congestion control models in depth. Another direction for future work is to investigate benefits of using RayNet in parallel OMNeT++ simulations where agents reside in multiple simulation partitions that run on different CPU cores or servers.

\bibliographystyle{ACM-Reference-Format}
\bibliography{main}


\begin{thebibliography}{63}


\ifx \showCODEN    \undefined \def \showCODEN     #1{\unskip}     \fi
\ifx \showDOI      \undefined \def \showDOI       #1{#1}\fi
\ifx \showISBNx    \undefined \def \showISBNx     #1{\unskip}     \fi
\ifx \showISBNxiii \undefined \def \showISBNxiii  #1{\unskip}     \fi
\ifx \showISSN     \undefined \def \showISSN      #1{\unskip}     \fi
\ifx \showLCCN     \undefined \def \showLCCN      #1{\unskip}     \fi
\ifx \shownote     \undefined \def \shownote      #1{#1}          \fi
\ifx \showarticletitle \undefined \def \showarticletitle #1{#1}   \fi
\ifx \showURL      \undefined \def \showURL       {\relax}        \fi
\providecommand\bibfield[2]{#2}
\providecommand\bibinfo[2]{#2}
\providecommand\natexlab[1]{#1}
\providecommand\showeprint[2][]{arXiv:#2}

\bibitem[Abbasloo et~al\mbox{.}(2020)]%
        {abbasloo2020classic}
\bibfield{author}{\bibinfo{person}{Soheil Abbasloo}, \bibinfo{person}{Chen-Yu Yen}, {and} \bibinfo{person}{H~Jonathan Chao}.} \bibinfo{year}{2020}\natexlab{}.
\newblock \showarticletitle{Classic meets modern: A pragmatic learning-based congestion control for the internet}. In \bibinfo{booktitle}{\emph{Proceedings of ACM SIGCOMM}}. \bibinfo{pages}{632--647}.
\newblock


\bibitem[Akyildiz et~al\mbox{.}(2001)]%
        {akyildiz2001tcp}
\bibfield{author}{\bibinfo{person}{Ian~F Akyildiz}, \bibinfo{person}{Giacomo Morabito}, {and} \bibinfo{person}{Sergio Palazzo}.} \bibinfo{year}{2001}\natexlab{}.
\newblock \showarticletitle{TCP-Peach: a new congestion control scheme for satellite IP networks}.
\newblock \bibinfo{journal}{\emph{IEEE/ACM Transactions on networking}} \bibinfo{volume}{9}, \bibinfo{number}{3} (\bibinfo{year}{2001}), \bibinfo{pages}{307--321}.
\newblock


\bibitem[Alizadeh et~al\mbox{.}(2010)]%
        {alizadeh2010data}
\bibfield{author}{\bibinfo{person}{Mohammad Alizadeh} {et~al\mbox{.}}} \bibinfo{year}{2010}\natexlab{}.
\newblock \showarticletitle{Data center {TCP (DCTCP)}}. In \bibinfo{booktitle}{\emph{Proceedings of ACM SIGCOMM}}. \bibinfo{pages}{63--74}.
\newblock


\bibitem[Aung et~al\mbox{.}(2023)]%
        {10070376}
\bibfield{author}{\bibinfo{person}{Nyothiri Aung} {et~al\mbox{.}}} \bibinfo{year}{2023}\natexlab{}.
\newblock \showarticletitle{VeSoNet: Traffic-Aware Content Caching for Vehicular Social Networks Using Deep Reinforcement Learning}.
\newblock \bibinfo{journal}{\emph{IEEE Transactions on Intelligent Transportation Systems}} \bibinfo{volume}{24}, \bibinfo{number}{8} (\bibinfo{year}{2023}), \bibinfo{pages}{8638--8649}.
\newblock


\bibitem[Barto et~al\mbox{.}(1983)]%
        {barto1983neuronlike}
\bibfield{author}{\bibinfo{person}{Andrew~G Barto}, \bibinfo{person}{Richard~S Sutton}, {and} \bibinfo{person}{Charles~W Anderson}.} \bibinfo{year}{1983}\natexlab{}.
\newblock \showarticletitle{Neuronlike adaptive elements that can solve difficult learning control problems}.
\newblock \bibinfo{journal}{\emph{IEEE Transactions on Systems, Man, and Cybernetics}} \bibinfo{number}{5} (\bibinfo{year}{1983}), \bibinfo{pages}{834--846}.
\newblock


\bibitem[Brakmo et~al\mbox{.}(1994)]%
        {brakmo1994tcp}
\bibfield{author}{\bibinfo{person}{Lawrence~S Brakmo}, \bibinfo{person}{Sean~W O'Malley}, {and} \bibinfo{person}{Larry~L Peterson}.} \bibinfo{year}{1994}\natexlab{}.
\newblock \showarticletitle{TCP Vegas: New techniques for congestion detection and avoidance}. In \bibinfo{booktitle}{\emph{Proceedings ofACM SIGCOMM}}. \bibinfo{pages}{24--35}.
\newblock


\bibitem[Brockman et~al\mbox{.}(2016)]%
        {1606.01540}
\bibfield{author}{\bibinfo{person}{Greg Brockman} {et~al\mbox{.}}} \bibinfo{year}{2016}\natexlab{}.
\newblock \bibinfo{title}{{OpenAI Gym}}.
\newblock
\newblock
\showeprint{arXiv:1606.01540}


\bibitem[Dong et~al\mbox{.}(2015)]%
        {dong2015pcc}
\bibfield{author}{\bibinfo{person}{Mo Dong}, \bibinfo{person}{Qingxi Li}, \bibinfo{person}{Doron Zarchy}, \bibinfo{person}{P~Brighten Godfrey}, {and} \bibinfo{person}{Michael Schapira}.} \bibinfo{year}{2015}\natexlab{}.
\newblock \showarticletitle{{PCC}: Re-architecting congestion control for consistent high performance}. In \bibinfo{booktitle}{\emph{Proceedings of USENIX NSDI}}. \bibinfo{pages}{395--408}.
\newblock


\bibitem[Dong et~al\mbox{.}(2018)]%
        {dong2018pcc}
\bibfield{author}{\bibinfo{person}{Mo Dong}, \bibinfo{person}{Tong Meng}, \bibinfo{person}{Doron Zarchy}, \bibinfo{person}{Engin Arslan}, \bibinfo{person}{Yossi Gilad}, \bibinfo{person}{Brighten Godfrey}, {and} \bibinfo{person}{Michael Schapira}.} \bibinfo{year}{2018}\natexlab{}.
\newblock \showarticletitle{{PCC} Vivace:{Online-Learning} Congestion Control}. In \bibinfo{booktitle}{\emph{Proceedings of USENIX NSDI}}. \bibinfo{pages}{343--356}.
\newblock


\bibitem[Dowling et~al\mbox{.}(2005)]%
        {1420665}
\bibfield{author}{\bibinfo{person}{J. Dowling}, \bibinfo{person}{E. Curran}, \bibinfo{person}{R. Cunningham}, {and} \bibinfo{person}{V. Cahill}.} \bibinfo{year}{2005}\natexlab{}.
\newblock \showarticletitle{Using feedback in collaborative reinforcement learning to adaptively optimize {MANET} routing}.
\newblock \bibinfo{journal}{\emph{IEEE Transactions on Systems, Man, and Cybernetics - Part A: Systems and Humans}} \bibinfo{volume}{35}, \bibinfo{number}{3} (\bibinfo{year}{2005}), \bibinfo{pages}{360--372}.
\newblock


\bibitem[Fuhrer et~al\mbox{.}(2022)]%
        {fuhrer2022implementing}
\bibfield{author}{\bibinfo{person}{Benjamin Fuhrer}, \bibinfo{person}{Yuval Shpigelman}, \bibinfo{person}{Chen Tessler}, \bibinfo{person}{Shie Mannor}, \bibinfo{person}{Gal Chechik}, \bibinfo{person}{Eitan Zahavi}, {and} \bibinfo{person}{Gal Dalal}.} \bibinfo{year}{2022}\natexlab{}.
\newblock \showarticletitle{Implementing Reinforcement Learning Datacenter Congestion Control in NVIDIA NICs}.
\newblock \bibinfo{journal}{\emph{arXiv preprint arXiv:2207.02295}} (\bibinfo{year}{2022}).
\newblock


\bibitem[Gaw{\l}owicz and Zubow(2019)]%
        {gawlowicz2019ns}
\bibfield{author}{\bibinfo{person}{Piotr Gaw{\l}owicz} {and} \bibinfo{person}{Anatolij Zubow}.} \bibinfo{year}{2019}\natexlab{}.
\newblock \showarticletitle{ns-3 meets {OpenAI Gym}: The playground for machine learning in networking research}. In \bibinfo{booktitle}{\emph{Proceedings of ACM MSWIM}}.
\newblock


\bibitem[Ha et~al\mbox{.}(2008)]%
        {ha2008cubic}
\bibfield{author}{\bibinfo{person}{Sangtae Ha}, \bibinfo{person}{Injong Rhee}, {and} \bibinfo{person}{Lisong Xu}.} \bibinfo{year}{2008}\natexlab{}.
\newblock \showarticletitle{CUBIC: a new TCP-friendly high-speed TCP variant}.
\newblock \bibinfo{journal}{\emph{ACM SIGOPS operating systems review}} \bibinfo{volume}{42}, \bibinfo{number}{5} (\bibinfo{year}{2008}), \bibinfo{pages}{64--74}.
\newblock


\bibitem[Haarnoja et~al\mbox{.}(2018)]%
        {haarnoja2018soft}
\bibfield{author}{\bibinfo{person}{Tuomas Haarnoja}, \bibinfo{person}{Aurick Zhou}, \bibinfo{person}{Pieter Abbeel}, {and} \bibinfo{person}{Sergey Levine}.} \bibinfo{year}{2018}\natexlab{}.
\newblock \showarticletitle{Soft actor-critic: Off-policy maximum entropy deep reinforcement learning with a stochastic actor}. In \bibinfo{booktitle}{\emph{Proceedings of ICML}}. \bibinfo{pages}{1861--1870}.
\newblock


\bibitem[He et~al\mbox{.}(2017)]%
        {he2017integrated}
\bibfield{author}{\bibinfo{person}{Ying He}, \bibinfo{person}{Nan Zhao}, {and} \bibinfo{person}{Hongxi Yin}.} \bibinfo{year}{2017}\natexlab{}.
\newblock \showarticletitle{Integrated networking, caching, and computing for connected vehicles: A deep reinforcement learning approach}.
\newblock \bibinfo{journal}{\emph{IEEE Transactions on Vehicular Technology}} \bibinfo{volume}{67}, \bibinfo{number}{1} (\bibinfo{year}{2017}), \bibinfo{pages}{44--55}.
\newblock


\bibitem[Horgan et~al\mbox{.}(2018)]%
        {horgan2018distributed}
\bibfield{author}{\bibinfo{person}{Dan Horgan} {et~al\mbox{.}}} \bibinfo{year}{2018}\natexlab{}.
\newblock \showarticletitle{Distributed prioritized experience replay}.
\newblock \bibinfo{journal}{\emph{arXiv preprint arXiv:1803.00933}} (\bibinfo{year}{2018}).
\newblock


\bibitem[Huang et~al\mbox{.}(2018)]%
        {huang2018qarc}
\bibfield{author}{\bibinfo{person}{Tianchi Huang}, \bibinfo{person}{Rui-Xiao Zhang}, \bibinfo{person}{Chao Zhou}, {and} \bibinfo{person}{Lifeng Sun}.} \bibinfo{year}{2018}\natexlab{}.
\newblock \showarticletitle{QARC: Video quality aware rate control for real-time video streaming based on deep reinforcement learning}. In \bibinfo{booktitle}{\emph{Proceedings of ACM Multimedia}}. \bibinfo{pages}{1208--1216}.
\newblock


\bibitem[Jay et~al\mbox{.}(2019)]%
        {jay2019deep}
\bibfield{author}{\bibinfo{person}{Nathan Jay}, \bibinfo{person}{Noga Rotman}, \bibinfo{person}{Brighten Godfrey}, \bibinfo{person}{Michael Schapira}, {and} \bibinfo{person}{Aviv Tamar}.} \bibinfo{year}{2019}\natexlab{}.
\newblock \showarticletitle{A deep reinforcement learning perspective on {Internet} congestion control}. In \bibinfo{booktitle}{\emph{Proceedings of ICML}}. \bibinfo{pages}{3050--3059}.
\newblock


\bibitem[Jiang et~al\mbox{.}(2019)]%
        {8629363}
\bibfield{author}{\bibinfo{person}{Wei Jiang}, \bibinfo{person}{Gang Feng}, \bibinfo{person}{Shuang Qin}, \bibinfo{person}{Tak Shing~Peter Yum}, {and} \bibinfo{person}{Guohong Cao}.} \bibinfo{year}{2019}\natexlab{}.
\newblock \showarticletitle{Multi-Agent Reinforcement Learning for Efficient Content Caching in Mobile D2D Networks}.
\newblock \bibinfo{journal}{\emph{IEEE Transactions on Wireless Communications}} \bibinfo{volume}{18}, \bibinfo{number}{3} (\bibinfo{year}{2019}), \bibinfo{pages}{1610--1622}.
\newblock


\bibitem[Kleinrock(2018)]%
        {kleinrock2018internet}
\bibfield{author}{\bibinfo{person}{Leonard Kleinrock}.} \bibinfo{year}{2018}\natexlab{}.
\newblock \showarticletitle{Internet congestion control using the power metric: Keep the pipe just full, but no fuller}.
\newblock \bibinfo{journal}{\emph{Ad hoc networks}}  \bibinfo{volume}{80} (\bibinfo{year}{2018}), \bibinfo{pages}{142--157}.
\newblock


\bibitem[Kliazovich et~al\mbox{.}(2006)]%
        {kliazovich2006tcp}
\bibfield{author}{\bibinfo{person}{Dzmitry Kliazovich}, \bibinfo{person}{Fabrizio Granelli}, {and} \bibinfo{person}{Daniele Miorandi}.} \bibinfo{year}{2006}\natexlab{}.
\newblock \showarticletitle{TCP Westwood+ enhancement in high-speed long-distance networks}. In \bibinfo{booktitle}{\emph{Proceedings of IEEE ICC}}. \bibinfo{pages}{710--715}.
\newblock


\bibitem[Kwon et~al\mbox{.}(2020)]%
        {9194445}
\bibfield{author}{\bibinfo{person}{Dohyun Kwon}, \bibinfo{person}{Joongheon Kim}, \bibinfo{person}{David~A. Mohaisen}, {and} \bibinfo{person}{Wonjun Lee}.} \bibinfo{year}{2020}\natexlab{}.
\newblock \showarticletitle{Self-adaptive power control with deep reinforcement learning for millimeter-wave Internet-of-vehicles video caching}.
\newblock \bibinfo{journal}{\emph{Journal of Communications and Networks}} \bibinfo{volume}{22}, \bibinfo{number}{4} (\bibinfo{year}{2020}), \bibinfo{pages}{326--337}.
\newblock


\bibitem[Lan et~al\mbox{.}(2019)]%
        {lan2019deep}
\bibfield{author}{\bibinfo{person}{Dehao Lan}, \bibinfo{person}{Xiaobin Tan}, \bibinfo{person}{Jinyang Lv}, \bibinfo{person}{Yang Jin}, {and} \bibinfo{person}{Jian Yang}.} \bibinfo{year}{2019}\natexlab{}.
\newblock \showarticletitle{A deep reinforcement learning based congestion control mechanism for NDN}. In \bibinfo{booktitle}{\emph{Proceedings of IEEE ICC}}. \bibinfo{pages}{1--7}.
\newblock


\bibitem[Liang et~al\mbox{.}(2018)]%
        {liang2018rllib}
\bibfield{author}{\bibinfo{person}{Eric Liang} {et~al\mbox{.}}} \bibinfo{year}{2018}\natexlab{}.
\newblock \showarticletitle{RLlib: Abstractions for distributed reinforcement learning}. In \bibinfo{booktitle}{\emph{Proceedings of ICML}}. \bibinfo{pages}{3053--3062}.
\newblock


\bibitem[Lillicrap et~al\mbox{.}(2015)]%
        {lillicrap2015continuous}
\bibfield{author}{\bibinfo{person}{Timothy~P Lillicrap} {et~al\mbox{.}}} \bibinfo{year}{2015}\natexlab{}.
\newblock \showarticletitle{Continuous control with deep reinforcement learning}.
\newblock \bibinfo{journal}{\emph{arXiv preprint arXiv:1509.02971}} (\bibinfo{year}{2015}).
\newblock


\bibitem[Mammeri(2019)]%
        {mammeri2019reinforcement}
\bibfield{author}{\bibinfo{person}{Zoubir Mammeri}.} \bibinfo{year}{2019}\natexlab{}.
\newblock \showarticletitle{Reinforcement learning based routing in networks: Review and classification of approaches}.
\newblock \bibinfo{journal}{\emph{IEEE Access}}  \bibinfo{volume}{7} (\bibinfo{year}{2019}), \bibinfo{pages}{55916--55950}.
\newblock


\bibitem[Mao et~al\mbox{.}(2017)]%
        {mao2017neural}
\bibfield{author}{\bibinfo{person}{Hongzi Mao}, \bibinfo{person}{Ravi Netravali}, {and} \bibinfo{person}{Mohammad Alizadeh}.} \bibinfo{year}{2017}\natexlab{}.
\newblock \showarticletitle{Neural adaptive video streaming with pensieve}. In \bibinfo{booktitle}{\emph{Proceedings of ACM SIGCOMM}}. \bibinfo{pages}{197--210}.
\newblock


\bibitem[Mascolo et~al\mbox{.}(2001)]%
        {mascolo2001tcp}
\bibfield{author}{\bibinfo{person}{Saverio Mascolo}, \bibinfo{person}{Claudio Casetti}, \bibinfo{person}{Mario Gerla}, \bibinfo{person}{Medy~Y Sanadidi}, {and} \bibinfo{person}{Ren Wang}.} \bibinfo{year}{2001}\natexlab{}.
\newblock \showarticletitle{TCP Westwood: Bandwidth estimation for enhanced transport over wireless links}. In \bibinfo{booktitle}{\emph{Proceedings of ACM MobiCom}}. \bibinfo{pages}{287--297}.
\newblock


\bibitem[Mastronarde and van~der Schaar(2011)]%
        {5986747}
\bibfield{author}{\bibinfo{person}{Nicholas Mastronarde} {and} \bibinfo{person}{Mihaela van~der Schaar}.} \bibinfo{year}{2011}\natexlab{}.
\newblock \showarticletitle{Fast Reinforcement Learning for Energy-Efficient Wireless Communication}.
\newblock \bibinfo{journal}{\emph{IEEE Transactions on Signal Processing}} \bibinfo{volume}{59}, \bibinfo{number}{12} (\bibinfo{year}{2011}), \bibinfo{pages}{6262--6266}.
\newblock


\bibitem[Mittal et~al\mbox{.}(2015)]%
        {mittal2015timely}
\bibfield{author}{\bibinfo{person}{Radhika Mittal} {et~al\mbox{.}}} \bibinfo{year}{2015}\natexlab{}.
\newblock \showarticletitle{TIMELY: RTT-based congestion control for the datacenter}.
\newblock \bibinfo{journal}{\emph{ACM SIGCOMM Computer Communication Review}} \bibinfo{volume}{45}, \bibinfo{number}{4} (\bibinfo{year}{2015}), \bibinfo{pages}{537--550}.
\newblock


\bibitem[Mnih et~al\mbox{.}(2013)]%
        {mnih2013playing}
\bibfield{author}{\bibinfo{person}{Volodymyr Mnih} {et~al\mbox{.}}} \bibinfo{year}{2013}\natexlab{}.
\newblock \showarticletitle{Playing {A}tari with deep reinforcement learning}.
\newblock \bibinfo{journal}{\emph{arXiv preprint arXiv:1312.5602}} (\bibinfo{year}{2013}).
\newblock


\bibitem[Moritz et~al\mbox{.}(2018)]%
        {moritz2018ray}
\bibfield{author}{\bibinfo{person}{Philipp Moritz} {et~al\mbox{.}}} \bibinfo{year}{2018}\natexlab{}.
\newblock \showarticletitle{Ray: A distributed framework for emerging {AI} applications}. In \bibinfo{booktitle}{\emph{Proceedings of USENIX OSDI}}. \bibinfo{pages}{561--577}.
\newblock


\bibitem[Naparstek and Cohen(2018)]%
        {naparstek2018deep}
\bibfield{author}{\bibinfo{person}{Oshri Naparstek} {and} \bibinfo{person}{Kobi Cohen}.} \bibinfo{year}{2018}\natexlab{}.
\newblock \showarticletitle{Deep multi-user reinforcement learning for distributed dynamic spectrum access}.
\newblock \bibinfo{journal}{\emph{IEEE Transactions on Wireless Communications}} \bibinfo{volume}{18}, \bibinfo{number}{1} (\bibinfo{year}{2018}), \bibinfo{pages}{310--323}.
\newblock


\bibitem[Nasehzadeh and Wang(2020)]%
        {9238811}
\bibfield{author}{\bibinfo{person}{Ali Nasehzadeh} {and} \bibinfo{person}{Ping Wang}.} \bibinfo{year}{2020}\natexlab{}.
\newblock \showarticletitle{A Deep Reinforcement Learning-Based Caching Strategy for Internet of Things}. In \bibinfo{booktitle}{\emph{Proceedings of IEEE/CIC ICCC}}. \bibinfo{pages}{969--974}.
\newblock


\bibitem[Netravali et~al\mbox{.}(2015)]%
        {netravali2015mahimahi}
\bibfield{author}{\bibinfo{person}{Ravi Netravali} {et~al\mbox{.}}} \bibinfo{year}{2015}\natexlab{}.
\newblock \showarticletitle{Mahimahi: Accurate {Record-and-Replay} for {HTTP}}. In \bibinfo{booktitle}{\emph{Proceedings of USENIX ATC}}. \bibinfo{pages}{417--429}.
\newblock


\bibitem[Nguyen and Reddi(2019)]%
        {nguyen2019deep}
\bibfield{author}{\bibinfo{person}{Thanh~Thi Nguyen} {and} \bibinfo{person}{Vijay~Janapa Reddi}.} \bibinfo{year}{2019}\natexlab{}.
\newblock \showarticletitle{Deep reinforcement learning for cyber security}.
\newblock \bibinfo{journal}{\emph{IEEE Transactions on Neural Networks and Learning Systems}} (\bibinfo{year}{2019}).
\newblock


\bibitem[Paasch and Bonaventure(2014)]%
        {paasch2014multipath}
\bibfield{author}{\bibinfo{person}{Christoph Paasch} {and} \bibinfo{person}{Olivier Bonaventure}.} \bibinfo{year}{2014}\natexlab{}.
\newblock \showarticletitle{Multipath {TCP}}.
\newblock \bibinfo{journal}{\emph{Commun. ACM}} \bibinfo{volume}{57}, \bibinfo{number}{4} (\bibinfo{year}{2014}), \bibinfo{pages}{51--57}.
\newblock


\bibitem[Qiao et~al\mbox{.}(2020)]%
        {8879573}
\bibfield{author}{\bibinfo{person}{Guanhua Qiao}, \bibinfo{person}{Supeng Leng}, \bibinfo{person}{Sabita Maharjan}, \bibinfo{person}{Yan Zhang}, {and} \bibinfo{person}{Nirwan Ansari}.} \bibinfo{year}{2020}\natexlab{}.
\newblock \showarticletitle{Deep Reinforcement Learning for Cooperative Content Caching in Vehicular Edge Computing and Networks}.
\newblock \bibinfo{journal}{\emph{IEEE Internet of Things Journal}} \bibinfo{volume}{7}, \bibinfo{number}{1} (\bibinfo{year}{2020}), \bibinfo{pages}{247--257}.
\newblock


\bibitem[Sacco et~al\mbox{.}(2021)]%
        {sacco2021owl}
\bibfield{author}{\bibinfo{person}{Alessio Sacco}, \bibinfo{person}{Matteo Flocco}, \bibinfo{person}{Flavio Esposito}, {and} \bibinfo{person}{Guido Marchetto}.} \bibinfo{year}{2021}\natexlab{}.
\newblock \showarticletitle{Owl: congestion control with partially invisible networks via reinforcement learning}. In \bibinfo{booktitle}{\emph{Proceedings of IEEE INFOCOM}}. \bibinfo{pages}{1--10}.
\newblock


\bibitem[Schulman et~al\mbox{.}(2017)]%
        {schulman2017proximal}
\bibfield{author}{\bibinfo{person}{John Schulman}, \bibinfo{person}{Filip Wolski}, \bibinfo{person}{Prafulla Dhariwal}, \bibinfo{person}{Alec Radford}, {and} \bibinfo{person}{Oleg Klimov}.} \bibinfo{year}{2017}\natexlab{}.
\newblock \showarticletitle{Proximal policy optimization algorithms}.
\newblock \bibinfo{journal}{\emph{arXiv preprint arXiv:1707.06347}} (\bibinfo{year}{2017}).
\newblock


\bibitem[Shimonishi et~al\mbox{.}(2005)]%
        {shimonishi2005improving}
\bibfield{author}{\bibinfo{person}{Hideyuki Shimonishi}, \bibinfo{person}{MY Sanadidi}, {and} \bibinfo{person}{Mario Gerla}.} \bibinfo{year}{2005}\natexlab{}.
\newblock \showarticletitle{Improving efficiency-friendliness tradeoffs of TCP in wired-wireless combined networks}. In \bibinfo{booktitle}{\emph{IEEE International Conference on Communications, 2005. ICC 2005. 2005}}, Vol.~\bibinfo{volume}{5}. IEEE, \bibinfo{pages}{3548--3552}.
\newblock


\bibitem[Sivaraman et~al\mbox{.}(2014)]%
        {sivaraman2014experimental}
\bibfield{author}{\bibinfo{person}{Anirudh Sivaraman}, \bibinfo{person}{Keith Winstein}, \bibinfo{person}{Pratiksha Thaker}, {and} \bibinfo{person}{Hari Balakrishnan}.} \bibinfo{year}{2014}\natexlab{}.
\newblock \showarticletitle{An experimental study of the learnability of congestion control}.
\newblock \bibinfo{journal}{\emph{ACM SIGCOMM Computer Communication Review}} \bibinfo{volume}{44}, \bibinfo{number}{4} (\bibinfo{year}{2014}), \bibinfo{pages}{479--490}.
\newblock


\bibitem[Song et~al\mbox{.}(2006)]%
        {song2006compound}
\bibfield{author}{\bibinfo{person}{Kun Tan~Jingmin Song}, \bibinfo{person}{Qian Zhang}, {and} \bibinfo{person}{Murari Sridharan}.} \bibinfo{year}{2006}\natexlab{}.
\newblock \showarticletitle{Compound TCP: A scalable and TCP-friendly congestion control for high-speed networks}.
\newblock \bibinfo{journal}{\emph{Proceedings of PFLDnet 2006}} (\bibinfo{year}{2006}).
\newblock


\bibitem[Stampa et~al\mbox{.}(2017)]%
        {stampa2017deep}
\bibfield{author}{\bibinfo{person}{Giorgio Stampa}, \bibinfo{person}{Marta Arias}, \bibinfo{person}{David S{\'a}nchez-Charles}, \bibinfo{person}{Victor Munt{\'e}s-Mulero}, {and} \bibinfo{person}{Albert Cabellos}.} \bibinfo{year}{2017}\natexlab{}.
\newblock \showarticletitle{A deep-reinforcement learning approach for software-defined networking routing optimization}.
\newblock \bibinfo{journal}{\emph{arXiv preprint arXiv:1709.07080}} (\bibinfo{year}{2017}).
\newblock


\bibitem[Sun et~al\mbox{.}(2019)]%
        {10.1145/3342280.3342317}
\bibfield{author}{\bibinfo{person}{Penghao Sun}, \bibinfo{person}{Junfei Li}, \bibinfo{person}{Zehua Guo}, \bibinfo{person}{Yang Xu}, \bibinfo{person}{Julong Lan}, {and} \bibinfo{person}{Yuxiang Hu}.} \bibinfo{year}{2019}\natexlab{}.
\newblock \showarticletitle{{SINET}: Enabling Scalable Network Routing with Deep Reinforcement Learning on Partial Nodes}. In \bibinfo{booktitle}{\emph{Proceedings of ACM SIGCOMM (Posters and Demos)}}. \bibinfo{pages}{88–89}.
\newblock


\bibitem[Sutton and Barto(2018)]%
        {sutton2018reinforcement}
\bibfield{author}{\bibinfo{person}{Richard~S Sutton} {and} \bibinfo{person}{Andrew~G Barto}.} \bibinfo{year}{2018}\natexlab{}.
\newblock \bibinfo{booktitle}{\emph{Reinforcement learning: An introduction}}.
\newblock \bibinfo{publisher}{MIT press}.
\newblock


\bibitem[Taleb et~al\mbox{.}(2006)]%
        {taleb2006refwa}
\bibfield{author}{\bibinfo{person}{Tarik Taleb}, \bibinfo{person}{Nei Kato}, {and} \bibinfo{person}{Yoshiaki Nemoto}.} \bibinfo{year}{2006}\natexlab{}.
\newblock \showarticletitle{REFWA: An efficient and fair congestion control scheme for LEO satellite networks}.
\newblock \bibinfo{journal}{\emph{IEEE/ACM Transactions on Networking}} \bibinfo{volume}{14}, \bibinfo{number}{5} (\bibinfo{year}{2006}), \bibinfo{pages}{1031--1044}.
\newblock


\bibitem[Tessler et~al\mbox{.}(2022)]%
        {tessler2022reinforcement}
\bibfield{author}{\bibinfo{person}{Chen Tessler}, \bibinfo{person}{Yuval Shpigelman}, \bibinfo{person}{Gal Dalal}, \bibinfo{person}{Amit Mandelbaum}, \bibinfo{person}{Doron~Haritan Kazakov}, \bibinfo{person}{Benjamin Fuhrer}, \bibinfo{person}{Gal Chechik}, {and} \bibinfo{person}{Shie Mannor}.} \bibinfo{year}{2022}\natexlab{}.
\newblock \showarticletitle{Reinforcement learning for datacenter congestion control}. In \bibinfo{booktitle}{\emph{Proceedings of the AAAI Conference on Artificial Intelligence}}, Vol.~\bibinfo{volume}{36}. \bibinfo{pages}{12615--12621}.
\newblock


\bibitem[Uprety and Rawat(2020)]%
        {uprety2020reinforcement}
\bibfield{author}{\bibinfo{person}{Aashma Uprety} {and} \bibinfo{person}{Danda~B Rawat}.} \bibinfo{year}{2020}\natexlab{}.
\newblock \showarticletitle{Reinforcement learning for iot security: A comprehensive survey}.
\newblock \bibinfo{journal}{\emph{IEEE Internet of Things Journal}} \bibinfo{volume}{8}, \bibinfo{number}{11} (\bibinfo{year}{2020}), \bibinfo{pages}{8693--8706}.
\newblock


\bibitem[Varga and Hornig(2008)]%
        {varga2008overview}
\bibfield{author}{\bibinfo{person}{Andr{\'a}s Varga} {and} \bibinfo{person}{Rudolf Hornig}.} \bibinfo{year}{2008}\natexlab{}.
\newblock \showarticletitle{An overview of the OMNeT++ simulation environment}. In \bibinfo{booktitle}{\emph{Proceedings of the 1st international conference on Simulation tools and techniques for communications, networks and systems \& workshops}}. \bibinfo{pages}{1--10}.
\newblock


\bibitem[Wang et~al\mbox{.}(2021)]%
        {9488770}
\bibfield{author}{\bibinfo{person}{Qi Wang}, \bibinfo{person}{Jianmin Liu}, \bibinfo{person}{Katia Jaffrès-Runser}, \bibinfo{person}{Yongqing Wang}, \bibinfo{person}{Chentao He}, \bibinfo{person}{Cunzhuang Liu}, {and} \bibinfo{person}{Yongjun Xu}.} \bibinfo{year}{2021}\natexlab{}.
\newblock \showarticletitle{INCdeep: Intelligent Network Coding with Deep Reinforcement Learning}. In \bibinfo{booktitle}{\emph{Proceedings of IEEE INFOCOM}}. \bibinfo{pages}{1--10}.
\newblock


\bibitem[Wang et~al\mbox{.}(2018)]%
        {wang2018deep}
\bibfield{author}{\bibinfo{person}{Shangxing Wang}, \bibinfo{person}{Hanpeng Liu}, \bibinfo{person}{Pedro~Henrique Gomes}, {and} \bibinfo{person}{Bhaskar Krishnamachari}.} \bibinfo{year}{2018}\natexlab{}.
\newblock \showarticletitle{Deep reinforcement learning for dynamic multichannel access in wireless networks}.
\newblock \bibinfo{journal}{\emph{IEEE Transactions on Cognitive Communications and Networking}} \bibinfo{volume}{4}, \bibinfo{number}{2} (\bibinfo{year}{2018}), \bibinfo{pages}{257--265}.
\newblock


\bibitem[Winstein and Balakrishnan(2013)]%
        {winstein2013tcp}
\bibfield{author}{\bibinfo{person}{Keith Winstein} {and} \bibinfo{person}{Hari Balakrishnan}.} \bibinfo{year}{2013}\natexlab{}.
\newblock \showarticletitle{Tcp ex machina: Computer-generated congestion control}.
\newblock \bibinfo{journal}{\emph{ACM SIGCOMM Computer Communication Review}} \bibinfo{volume}{43}, \bibinfo{number}{4} (\bibinfo{year}{2013}), \bibinfo{pages}{123--134}.
\newblock


\bibitem[Wu et~al\mbox{.}(2013)]%
        {wu2010ictcp}
\bibfield{author}{\bibinfo{person}{Haitao Wu}, \bibinfo{person}{Zhenqian Feng}, \bibinfo{person}{Chuanxiong Guo}, {and} \bibinfo{person}{Yongguang Zhang}.} \bibinfo{year}{2013}\natexlab{}.
\newblock \showarticletitle{{ICTCP: Incast Congestion Control for TCP in Data-Center Networks}}.
\newblock \bibinfo{journal}{\emph{IEEE/ACM Transactions on Networking}} \bibinfo{volume}{21}, \bibinfo{number}{2} (\bibinfo{year}{2013}), \bibinfo{pages}{345--358}.
\newblock


\bibitem[Xu et~al\mbox{.}(2020)]%
        {10.1145/3424978.3425004}
\bibfield{author}{\bibinfo{person}{Chunlei Xu}, \bibinfo{person}{Weijin Zhuang}, {and} \bibinfo{person}{Hong Zhang}.} \bibinfo{year}{2020}\natexlab{}.
\newblock \showarticletitle{A Deep-Reinforcement Learning Approach for SDN Routing Optimization}. In \bibinfo{booktitle}{\emph{Proceedings of CSAE}}.
\newblock


\bibitem[Xu et~al\mbox{.}(2019)]%
        {xu2019experience}
\bibfield{author}{\bibinfo{person}{Zhiyuan Xu}, \bibinfo{person}{Jian Tang}, \bibinfo{person}{Chengxiang Yin}, \bibinfo{person}{Yanzhi Wang}, {and} \bibinfo{person}{Guoliang Xue}.} \bibinfo{year}{2019}\natexlab{}.
\newblock \showarticletitle{Experience-driven congestion control: When multi-path TCP meets deep reinforcement learning}.
\newblock \bibinfo{journal}{\emph{IEEE Journal on Selected Areas in Communications}} \bibinfo{volume}{37}, \bibinfo{number}{6} (\bibinfo{year}{2019}), \bibinfo{pages}{1325--1336}.
\newblock


\bibitem[Xu et~al\mbox{.}(2021)]%
        {9274515}
\bibfield{author}{\bibinfo{person}{Zhiyuan Xu}, \bibinfo{person}{Dejun Yang}, \bibinfo{person}{Jian Tang}, \bibinfo{person}{Yinan Tang}, \bibinfo{person}{Tongtong Yuan}, \bibinfo{person}{Yanzhi Wang}, {and} \bibinfo{person}{Guoliang Xue}.} \bibinfo{year}{2021}\natexlab{}.
\newblock \showarticletitle{An Actor-Critic-Based Transfer Learning Framework for Experience-Driven Networking}.
\newblock \bibinfo{journal}{\emph{IEEE/ACM Transactions on Networking}} \bibinfo{volume}{29}, \bibinfo{number}{1} (\bibinfo{year}{2021}), \bibinfo{pages}{360--371}.
\newblock


\bibitem[Yu et~al\mbox{.}(2018)]%
        {8502806}
\bibfield{author}{\bibinfo{person}{Changhe Yu}, \bibinfo{person}{Julong Lan}, \bibinfo{person}{Zehua Guo}, {and} \bibinfo{person}{Yuxiang Hu}.} \bibinfo{year}{2018}\natexlab{}.
\newblock \showarticletitle{DROM: Optimizing the Routing in Software-Defined Networks With Deep Reinforcement Learning}.
\newblock \bibinfo{journal}{\emph{IEEE Access}}  \bibinfo{volume}{6} (\bibinfo{year}{2018}), \bibinfo{pages}{64533--64539}.
\newblock


\bibitem[Zhao et~al\mbox{.}(2021)]%
        {9141401}
\bibfield{author}{\bibinfo{person}{Jiuxia Zhao}, \bibinfo{person}{Minjia Mao}, \bibinfo{person}{Xi Zhao}, {and} \bibinfo{person}{Jianhua Zou}.} \bibinfo{year}{2021}\natexlab{}.
\newblock \showarticletitle{A Hybrid of Deep Reinforcement Learning and Local Search for the Vehicle Routing Problems}.
\newblock \bibinfo{journal}{\emph{IEEE Transactions on Intelligent Transportation Systems}} \bibinfo{volume}{22}, \bibinfo{number}{11} (\bibinfo{year}{2021}), \bibinfo{pages}{7208--7218}.
\newblock


\bibitem[Zhao et~al\mbox{.}(2019)]%
        {8703471}
\bibfield{author}{\bibinfo{person}{Lei Zhao}, \bibinfo{person}{Jiadai Wang}, \bibinfo{person}{Jiajia Liu}, {and} \bibinfo{person}{Nei Kato}.} \bibinfo{year}{2019}\natexlab{}.
\newblock \showarticletitle{Routing for Crowd Management in Smart Cities: A Deep Reinforcement Learning Perspective}.
\newblock \bibinfo{journal}{\emph{IEEE Communications Magazine}} \bibinfo{volume}{57}, \bibinfo{number}{4} (\bibinfo{year}{2019}), \bibinfo{pages}{88--93}.
\newblock


\bibitem[Zhong et~al\mbox{.}(2020)]%
        {8964499}
\bibfield{author}{\bibinfo{person}{Chen Zhong}, \bibinfo{person}{M.~Cenk Gursoy}, {and} \bibinfo{person}{Senem Velipasalar}.} \bibinfo{year}{2020}\natexlab{}.
\newblock \showarticletitle{Deep Reinforcement Learning-Based Edge Caching in Wireless Networks}.
\newblock \bibinfo{journal}{\emph{IEEE Transactions on Cognitive Communications and Networking}} \bibinfo{volume}{6}, \bibinfo{number}{1} (\bibinfo{year}{2020}), \bibinfo{pages}{48--61}.
\newblock


\bibitem[Zhu et~al\mbox{.}(2018)]%
        {zhu2018deep}
\bibfield{author}{\bibinfo{person}{Hao Zhu}, \bibinfo{person}{Yang Cao}, \bibinfo{person}{Wei Wang}, \bibinfo{person}{Tao Jiang}, {and} \bibinfo{person}{Shi Jin}.} \bibinfo{year}{2018}\natexlab{}.
\newblock \showarticletitle{Deep reinforcement learning for mobile edge caching: Review, new features, and open issues}.
\newblock \bibinfo{journal}{\emph{IEEE Network}} \bibinfo{volume}{32}, \bibinfo{number}{6} (\bibinfo{year}{2018}), \bibinfo{pages}{50--57}.
\newblock


\bibitem[Zhu et~al\mbox{.}(2019)]%
        {8542696}
\bibfield{author}{\bibinfo{person}{Hao Zhu}, \bibinfo{person}{Yang Cao}, \bibinfo{person}{Xiao Wei}, \bibinfo{person}{Wei Wang}, \bibinfo{person}{Tao Jiang}, {and} \bibinfo{person}{Shi Jin}.} \bibinfo{year}{2019}\natexlab{}.
\newblock \showarticletitle{Caching Transient Data for Internet of Things: A Deep Reinforcement Learning Approach}.
\newblock \bibinfo{journal}{\emph{IEEE Internet of Things Journal}} \bibinfo{volume}{6}, \bibinfo{number}{2} (\bibinfo{year}{2019}), \bibinfo{pages}{2074--2083}.
\newblock


\end{thebibliography}

\end{document}